%% file: main.tex
\documentclass[ijoc,nonblindrev]{informs4}

\usepackage{amsmath,amssymb}
\usepackage{booktabs}
\usepackage{graphicx}
\usepackage{url}

\OneAndAHalfSpacedXI
\TheoremsNumberedThrough
\ECRepeatTheorems
\EquationsNumberedThrough

\input{numbers}

\newcommand{\cA}{\mathcal{A}}
\newcommand{\cB}{\mathcal{B}}
\newcommand{\cC}{\mathcal{C}}

\newcommand{\cT}{\mathcal{T}}
\newcommand{\cX}{\mathcal{X}}
\newcommand{\R}{\mathbb{R}}
\newcommand{\Z}{\mathbb{Z}}
\newcommand{\Rbar}{\overline{\R}}
\newcommand{\vfun}{v}

\newcommand{\Pib}{\widehat{\Pi}}

\begin{document}

\RUNAUTHOR{Li and Hai}
\RUNTITLE{Falsification-Based Verification of LLM-Generated Optimization Models}

\TITLE{Falsification-Based Verification of LLM-Generated Optimization Models: Sound Test Batteries and Their Detection Limits}

\ARTICLEAUTHORS{%
\AUTHOR{Haifeng Li}
\AFF{School of Information, Central University of Finance and Economics,
\EMAIL{mydlhf@cufe.edu.cn}}
\AUTHOR{Mo Hai}
\AFF{School of Information, Central University of Finance and Economics,
corresponding author, \EMAIL{haimo@cufe.edu.cn}}
}

\ABSTRACT{%
Large language models now translate natural-language descriptions of
decision problems into solver-ready optimization models, and they fail
silently.  A generated model often runs and still encodes the wrong
problem, while standard evaluation compares optimal values against
labeled answers that deployment does not provide.  How to certify such
a model without any reference is the question this paper addresses.  We
develop falsification-based verification.  Every numeric quantity in a
problem description plays a role that the text itself states, such as a
capacity, a requirement, or a unit cost, and any correct model must
respond to changes in these quantities as the stated roles dictate.
From duality and sensitivity analysis we derive a battery of
solver-based tests that are individually sound, so a violation
certifies a faulty model and the false-positive rate is zero by design.
We characterize the errors that no test of this kind can see, give
conditions under which each canonical error class is detected with
certainty, and prove that perturbation testers with tuned thresholds
cannot be simultaneously sound and nontrivial.  Across
\numSeedsTotal{} ground-truth models, a synthetic family, and four
public benchmarks with two generators, the battery flags
\numCleanFPRpct{} of faithful models while a threshold tester flags
\numReloopFPRpct{}; it detects \numCoreDetpct{} of core formulation
errors, \numCondDetpct{} under certified preconditions, and
\numBlindDetpct{} of the errors that value-based scoring provably
cannot see, and it reproduces the predicted detectability pattern
including its blind spots.  Every flag carries a machine-checkable
certificate that localizes the defect, and a full audit costs about
\numAvgSolves{} millisecond-scale solver calls per model.  Classical
sensitivity analysis and duality thus offer a rigorous, label-free
audit that complements existing evaluation of AI-generated
optimization models.%
}


\maketitle

\section{Introduction}\label{sec:intro}

An operations researcher who receives a model from a junior colleague does
not check it by re-deriving it from scratch.  She interrogates it.  If we
double the budget, does the plan get better?  If demand were zero, would we
really spend nothing?  Why does adding a truck make things worse?  These
questions require no knowledge of the correct model.  They exploit the fact
that the \emph{structure} of the problem dictates how any correct
formulation must respond to changes in its data.  A model that spends money
at zero demand, or deteriorates when given more capacity, convicts itself.

This paper turns that interrogation into mathematics and software for
optimization models generated by large language models, the setting
where it has become the practical bottleneck.  A rapidly
growing line of work fine-tunes or prompts LLMs to translate natural-language
problem descriptions into solver-ready mathematical programs
\citep{ramamonjison2022nl4opt,xiao2024chainofexperts,ahmaditeshnizi2024optimus,tang2024orlm,jiang2025llmopt,lu2025optmath}.
The reported progress is real, but it is measured almost exclusively by
execution accuracy, which generates a model, runs a solver, and compares the
optimal value against a labeled reference.  Deployment offers no such
reference.  Worse, execution accuracy is blind in both directions.  A
generated model can return the labeled value by accident while encoding the
wrong feasible set, and a correct model can be scored wrong because the label
itself is erroneous.  \citet{chen2025solverinformed} found it necessary
to review and correct the
NL4OPT and IndustryOR labels before evaluating on them.  The gap
between solver feasibility and semantic correctness reaches up to ninety
percentage points on compositional problems \citep{lian2026reloop}.
Silent failure dominates syntax error as the failure mode in practice,
and it is the one failure mode that current evaluation cannot see.

The community has responded with three families of safeguards.
\emph{Reference-based evaluation} compares generated models against
human-written ground truth \citep{wang2025orgeval}; it is rigorous but
presupposes the artifact whose absence defines the problem.
\emph{LLM-as-verifier} schemes ask a second model to critique the first
\citep{fang2026trival,zhou2026optverifier,li2026optargus}; they inherit the
reviewers' blind spots and offer no guarantees, since self-critique is known
to be unreliable just when the generator is unreliable
\citep{huang2024llmcannotselfcorrect}.  Most recently, and closest to this
paper, \emph{behavioral testing} has emerged: \citet{lian2026reloop} perturb
a parameter of the generated code toward an extreme, re-solve, and flag the
model when the objective moves by less than a tuned threshold of five
percent, interpreting insensitivity as a missing constraint.  The
instinct to route verification through the solver is right, and
\citet{lian2026reloop} demonstrate that it repairs real failures.
What this line of work still lacks is its theory, in three respects.
First, the thresholds are heuristic.  A correct model whose capacity is
slack at the probe fails the test, so the false-positive rate is
structurally bounded away from zero, and false alarms trigger repairs that
damage correct models.  Second, the tests ask only whether the objective
reacts at all, leaving unused the sharper information that optimization
theory supplies about the direction, the curvature, the limit, and the
symmetry of the reaction.  Third, and most fundamentally, little is known
about what such testing can and cannot detect; a soundness statement, a
characterization of invisible errors, and a separation from the naive
alternatives have yet to be given.  Behavioral testing has, in effect,
begun to reuse sensitivity analysis as a testing device, without yet
drawing on the fifty years of theory that make it trustworthy.

\subsection{This paper}

We supply that theory, and we show that it pays measurable dividends.  Our
starting point is a change of object.  A word problem does not describe one
optimization instance; it describes a parametric family.  Every number in
the text is a typed quantity that we call a slot, and its semantic role as a
capacity, a requirement, a per-unit rate, a cost, a reward, or a share of a
total is asserted by the text itself.  A candidate model is faithful only if it
responds to slot changes the way the asserted roles dictate.  Faithfulness therefore becomes testable by relations alone.  We transform
slot values, re-solve the candidate, and compare the observed response
against necessary conditions derived from optimization theory.  No ground-truth model, label, or reference
value is consulted at any point.

The result is a \emph{sound}, oracle-free verifier for LLM-generated
LPs and MILPs, a verifier whose alarms are theorems, together with a
provable map of what no such verifier can see.  The paper makes four
contributions.

\emph{(i) A sound battery.}  Section~\ref{sec:battery} derives six classes
of structure tests.  Directional tests check comparative statics in
right-hand sides, objective coefficients, and technology rates; curvature
tests check midpoint convexity of the linear-programming value function;
certified crush probes drive a resource to zero or a requirement toward
infinity and demand infeasibility, with no threshold anywhere;
prohibitive-limit probes price an activity out of the model and require
exact agreement with the optimum under a hard zero, which a polyhedral
finite-threshold argument justifies and a two-scale convergence criterion
implements; annihilation probes zero all requirements and demand exactly
zero cost; exchange probes swap the complete slot bundles of
interchangeable entities and demand an unchanged value.  Every test is
proved \emph{sound} for the asserted
roles (Theorem~\ref{thm:soundness}).  A faithful candidate passes all tests,
under any degeneracy, so a violation is a certificate of
unfaithfulness and the battery's false-positive rate is zero by
construction, a property that threshold-based testing cannot have
(Proposition~\ref{prop:threshold}).  A second, assertion-free layer of
coherence tests (homogeneity, relaxation dominance, removal monotonicity,
dual subgradient inequalities) is sound for \emph{any} mixed-integer linear
program and catches computational and reporting corruption in generated code.

\emph{(ii) Detection limits.}  Section~\ref{sec:limits} characterizes the
blind set.  Uniform positive rescaling of the objective is invisible to every
oracle-free relational battery, and it is provably harmless because it
preserves the optimal decisions (Proposition~\ref{prop:gauge}).
Constraints that never bind on the probe orbit are invisible relative to that
orbit (Proposition~\ref{prop:redundant}).  Linear misuse of a single slot, the classic unit error in which one
coefficient is off by a consistent factor, is invisible to all single-slot
probes but is exposed by exchange probes whenever the slot lies in a
certified interchangeable bundle (Proposition~\ref{prop:reparam}).  We further prove an asymmetry.  The
prohibitive-limit equality holds on the cost side of minimization problems
but degrades to a one-sided inequality for reward annihilation in
maximization problems (Proposition~\ref{prop:asymmetry}).  Conversely,
Theorem~\ref{thm:detect} gives sufficient conditions under which each
canonical error class is detected with certainty by a specific test class.
The classes cover constraint direction flips, constraint omission,
right-hand-side misassignment, dropped objective terms, wrong optimization
sense, entity misbinding, and hard-coded data.  Together the two results yield a falsifiable prediction in the form of a
class-by-error detectability matrix, which our experiments reproduce,
including its zeros.

\emph{(iii) A verifiable modeling interface.}  Section~\ref{sec:vmi}
operationalizes the theory for free-form generation.  A first LLM pass
extracts the typed slot table from the text; a second pass generates the
model in a JSON intermediate representation that may reference quantities only
through slot identifiers.  Data binding, the map from text quantities to
model coefficients that behavioral testing silently presupposes, thereby
becomes explicit and checkable, and the battery applies to any model, from
any generator, without touching solver code.

\emph{(iv) Computational evidence.}  Section~\ref{sec:experiments}
evaluates the theory on \numSeedsTotal{} ground-truth models derived
from the NL4OPT annotations, a controlled synthetic family, and four
benchmark families generated end to end by two generators, a locally
hosted Qwen2.5-7B and DeepSeek-V4-flash.  The battery flags
\numCleanFPRpct{} of faithful models; the reimplemented threshold
tester of \citet{lian2026reloop} flags \numReloopFPRpct{} of the same
models.  Detection matches the theory class by class, including the
predicted zeros of the invisibility controls, and certified
preconditions raise it from \numUncondDetpct{} to \numCondDetpct{} on
the conditional classes.  Among the mutants whose optimal value
coincides with the correct one, which execution accuracy cannot catch
even with perfect labels, the battery still convicts \numBlindDetpct{}.
Deployment casts the battery as an auditor.  As a selection filter it
is cost-free yet statistically indistinguishable from majority voting;
as an auditor, every flag carries a certificate, adjudication of every
value-correct flagged candidate traces \numAdjDeservedpct{} of the
flags to genuine structural defects, and a binding-consistency
refinement removes most of the remaining extraction noise while
preserving soundness.  A full audit costs \numAvgSolves{}
millisecond-scale LP solves per model, and label and annotation defects
in the public benchmarks surface as a by-product.

\subsection{Why this matters for operations research}

The verification question is a central adoption bottleneck for
automated modeling, and many of the tools for answering it come from
operations research.  The necessary conditions we test, monotonicity of
the value function in resources, convexity in right-hand sides,
shadow-price consistency, symmetry under relabeling, and exactness of
prohibitive limits, form a stock of provably valid test relations that
sensitivity analysis and duality theory have accumulated and that
generic program synthesis lacks (Section~\ref{sec:related}).  Using
this capital as a verification technology for the AI systems the field
is beginning to deploy requires knowing its limits, which we map.

\section{Related Work}\label{sec:related}

\subsection{LLMs for optimization modeling}
Translating natural language to mathematical programs began as a semantic
parsing task with the NL4OPT competition \citep{ramamonjison2022nl4opt} and
has since split into prompting-based agents
\citep{xiao2024chainofexperts,ahmaditeshnizi2024optimus,mostajabdaveh2024multiagent},
fine-tuned open models \citep{tang2024orlm,jiang2025llmopt,lu2025optmath,yang2025optibench},
reinforcement learning with solver-execution rewards
\citep{chen2025solverinformed,guan2026evom,luo2026vrpcoder}, and inference
pipelines with structured reasoning \citep{zhao2026strategist,zhaok2026miniopt}.
Benchmarks have multiplied alongside, from NL4OPT, MAMO
\citep{huang2024mamo}, IndustryOR \citep{tang2024orlm}, OptMATH
\citep{lu2025optmath}, and OptiBench \citep{yang2025optibench} to
compositional suites such as RetailOpt-190 \citep{lian2026reloop} and
ConstraintBench \citep{feng2026constraintbench}.  Two facts from this
literature motivate us.  The same model that reaches roughly 85\% on
textbook-style problems drops to 38\% on industrial ones
\citep{tang2024orlm}, and the benchmark labels themselves are documented
to be unreliable \citep{chen2025solverinformed}.  Both facts locate the
binding constraint in the ability to tell, for a specific generated
model with no reference available, whether it is right.

\subsection{Verification of generated models}
Approaches divide by what they trust.  \emph{Reference-based} evaluation
compares against a human ground-truth model, as in the graph-isomorphism
test of ORGEval \citep{wang2025orgeval} and the conversion audits of
\citet{klamkin2025dualschool}, and therefore needs the human artifact we
lack.  \emph{LLM-trusting} schemes validate specification, formulation, and
code with LLMs (TriVAL, \citealt{fang2026trival}), cross-examine
structure and solutions (Opt-Verifier, \citealt{zhou2026optverifier}),
coordinate specialist auditors under a hallucination taxonomy
(OptArgus, \citealt{li2026optargus}), diagnose infeasible models
conversationally (OptiChat, \citealt{chen2025optichat}), or let agent
frameworks generate their own test code
\citep{zadorojniy2025validation}.  All ultimately ask a language model
to certify a language model.  \emph{Solver-trusting}
methods began with ReLoop \citep{lian2026reloop}, which introduces
constraint-presence and objective-presence testing by extreme
perturbation with graduated thresholds; OptiLoop \citep{xu2026optiloop}
extends the idea to coordination protocols, and VRPCoder
\citep{luo2026vrpcoder} verifies vehicle-routing programs by injecting
feasible and one-violation probe solutions and reuses the verifier as a
training reward.  Our work belongs to this third family and supplies
the theory it lacks, with a formal framework, soundness guarantees in
place of thresholds, a test taxonomy generalizing presence checks,
detection limits, and a data-binding interface that makes the scheme
well-defined for arbitrary generators.

\subsection{Metamorphic testing and the oracle problem}
Software engineering addresses testing without expected outputs through
metamorphic relations among inputs and outputs
\citep{chen2018metamorphic,segura2016survey}; LLM-generated code has been
validated by cross-prompt consistency \citep{xia2024metamorphicprompt}, and
surveys catalog relation-generation strategies
\citep{li2024mrgeneration}.  The perennial weakness is the source of
relations, which are guessed, mined, or crowd-sourced, so that their
validity is itself uncertain.  Our observation is that optimization is the rare domain
where sensitivity analysis, duality, and
symmetry supply an inexhaustible stock of provably valid metamorphic
relations, and where a violation means a certificate of unfaithfulness
relative to typed assertions.  Mutation analysis, which we use to measure detection power,
follows the classical methodology \citep{jia2011mutation} and the
optimization-specific mutation study of \citet{zadorojniy2025validation},
whose binding-constraint counting bound we strengthen to a
class-by-error characterization with matched impossibility results.

\subsection{Sensitivity analysis and comparative statics}
The structural results we repurpose are classical, covering
value-function convexity and shadow-price subgradients in linear
programming \citep{bertsimas1997introduction,vanderbei2020linear},
monotone comparative statics \citep{topkis1998supermodularity}, and Le
Chatelier-type principles \citep{milgrom1996lechatelier}.  Their
traditional role is interpretation, what-if analysis for a model
assumed correct.  Following
the inversion first exploited heuristically by \citet{lian2026reloop}, we
use them as falsifiers for models assumed suspect; our contribution is to
carry out the inversion with the theory's own standards of rigor, including
its impossibility half.  The exchange tests connect to symmetry detection in
integer programming \citep{margot2010symmetry}, and the certified crush
probes to irreducible infeasible subsystem analysis \citep{chinneck2008feasibility}.

\subsection{Trustworthy AI pipelines in operations}
This paper joins a broader agenda of pairing learned components with
optimization-theoretic guarantees at deployment time, as in constraint
learning \citep{fajemisin2024constraintlearning} and decision-focused
learning under verification \citep{sadana2025contextual}.  Sound
verification is the entry gate through which auto-formulated models
must pass to be usable in regulated and high-stakes operations.

\section{The Verification Problem}\label{sec:framework}

\subsection{Modeling tasks as parametric families}

A \emph{modeling task} is a natural-language description $T$ containing $p$
numeric quantities $\theta^0=(\theta^0_1,\dots,\theta^0_p)\in\R^p$.  We call
the positions $j=1,\dots,p$ \emph{slots}.  The text asserts, for each slot, a
semantic type; we use the vocabulary
\[
\texttt{qtype}(j)\in\{\textsf{capacity},\ \textsf{requirement},\
\textsf{rate},\ \textsf{cost},\ \textsf{reward},\ \textsf{ratio}\},
\]
optionally an \emph{entity} $e(j)$ (the activity or product the quantity
belongs to), and a global \emph{sense} $\sigma^\ast\in\{\min,\max\}$.  We
write $\cA=(\sigma^\ast,\texttt{qtype},e)$ for this \emph{assertion set}.
Assertions are cheap because they restate the surface of the text and
involve no modeling.  A phrase such as ``at most 200'' marks a capacity,
and ``\$10 per trip'' for vans marks a per-unit cost of the van
activity.  In our annotated experiments they come mechanically from the
NL4OPT ground-truth declarations; in deployment they come from a dedicated
extraction pass (Section~\ref{sec:vmi}).

A \emph{candidate model} is a parametric mixed-integer linear program
\begin{equation}\label{eq:candidate}
\Pib:\quad \theta\ \mapsto\
\Big[\ \underset{x\in\cX(\theta)}{\sigma}\ \ f(x;\theta)
= c(\theta)^\top x + c_0(\theta)\ \Big],
\qquad
\cX(\theta)=\{x\in\R^n_{\ge0}:\ A(\theta)x \lessgtr b(\theta),\ x_I\in\Z\},
\end{equation}
where $\sigma\in\{\min,\max\}$, $\lessgtr$ denotes row-wise senses in
$\{\le,\ge,=\}$, and $I$ indexes integer variables.  The candidate is
produced by an untrusted generator; we may execute it (solve at any
$\theta$) but not trust it.  Define the \emph{value function}
$\vfun(\theta)\in\Rbar$ of $\Pib$ with the conventions
$\vfun=+\infty$ (min) or $-\infty$ (max) if $\cX(\theta)=\emptyset$, and the
opposite infinities for unboundedness.

\begin{definition}[Role-faithfulness]\label{def:faithful}
Candidate \eqref{eq:candidate} is \emph{role-faithful} to $\cA$ if
$\sigma=\sigma^\ast$ and there is a representation of
$(c,c_0,A,b)$ as functions of $\theta$ in which each slot $j$ enters
\emph{only} at its asserted position, as the identity map:
a \textsf{capacity} slot appears only as the right-hand side of one row of
sense $\le$ (a \textsf{requirement} slot: of sense $\ge$); a \textsf{rate}
slot with entity $e$ appears only as the coefficient of $e$'s variable in one
constraint row; a \textsf{cost}/\textsf{reward} slot with entity $e$ only as
the objective coefficient of $e$'s variable; a \textsf{ratio} slot $\rho$
with entity $e$ only through a row expressing that $e$'s activity is at
least/at most a $\rho$-share of the relevant total.  All remaining data of
$\Pib$ are constants in $\theta$.
\end{definition}

\begin{definition}[Structure- and full faithfulness]\label{def:faithful2}
When a complete declared formulation $S^\ast$ is available (annotated mode),
$\Pib$ is \emph{structure-faithful} if $\cX_{\Pib}(\theta)=\cX_{S^\ast}(\theta)$
for all $\theta$ in a neighborhood of the probe set (the orbit
$\Theta_{\cB}$ of Section~\ref{sec:limits}), and \emph{faithful} if in
addition its objective agrees with $S^\ast$'s up to the gauge group of
Proposition~\ref{prop:gauge}.
\end{definition}

\subsection{Oracle-free tests}

A \emph{probe} is a finite collection of solver calls on transformed
instances: slot substitutions $\theta^0\mapsto\tau(\theta^0)$ (change one or
several slot values), model-level operations (relax integrality; fix an
entity's variables to zero), or both.  A \emph{test} $t=(\pi,\varphi)$ pairs
a probe $\pi$ with a predicate $\varphi$ over the observed tuple of statuses
and values (and, for Layer B, duals and incumbents).  A \emph{battery}
$\cB$ is a finite set of tests; the verifier flags $\Pib$ if some test fails.

\begin{definition}[Soundness and oracle-freeness]\label{def:sound}
A test is \emph{sound} for a class $\cC$ of candidates if every
$\Pib\in\cC$ passes it (in exact arithmetic).  A battery is sound for $\cC$
if all its tests are.  A test is \emph{oracle-free} if $\varphi$ does not
depend on any reference model, reference optimal value, or human label.
\end{definition}

Soundness separates certification from anomaly detection.  When a sound
battery fires, the candidate is provably not in $\cC$, and the failed test
is a certificate of unfaithfulness.  Silence is never claimed to certify
correctness.  Verification is thus falsification in the Popperian
sense, with the class $\cC$ of faithful candidates playing the
role of the theory under test.  Two design rules keep the implemented
battery sound in floating point.  Predicates use one-sided tolerances, so
that a violation must exceed $\varepsilon$ in the forbidden direction,
and every test carries a precondition that is certified before the test
may fire, on the declared structure in annotated mode and on the
candidate's own structure in deployment mode.  For a faithful candidate the two coincide, and a
certification computed on an unfaithful candidate can only affect power,
never soundness.

\subsection{What the verifier reports}

For a candidate $\Pib$ the verifier returns the list of failed tests with
their probes.  A typical report states that raising the budget slot from 200
to 401 worsened the objective from 3{,}540 to 3{,}890, which localizes the
misbehavior to one slot and one structural property.  We evaluate the
\emph{detector} and leave repair policies, whether to regenerate, patch,
or escalate to a human, as downstream choices; misdirected repair
driven by unsound flags is the failure mode our soundness guarantee
eliminates.

\section{A Sound Battery from Duality and Comparative Statics}\label{sec:battery}

Table~\ref{tab:battery} summarizes the battery.  Layer A tests structure
against assertions.  Layer B tests computational coherence and needs no
assertions.  We state all results in the main text and give complete proofs
in E-Companion~\ref{ec:proofs}.

\begin{table}[t]
\TABLE
{The falsification battery.  Each row is a family of tests instantiated per
slot, row, or entity.  The precondition column is certified before the test fires;
$\vfun$ is the candidate's value function.\label{tab:battery}}
{\begin{tabular}{@{}lllll@{}}
\toprule
Class & Probe & Predicate (asserted sense) & Precondition & Basis \\
\midrule
A0 static & none & declared sense $=\sigma^\ast$ & --- & assertion \\
A1 rhs direction & capacity/requirement slot $\pm$ & $\vfun$ weakly improves/worsens & --- & set inclusion \\
A2 coef direction & cost/reward/rate/ratio slot $+$ & $\vfun$ moves weakly as asserted & $x\ge0$ & pointwise monotonicity \\
A3 rhs curvature & rhs slot at $b\!-\!\delta,b,b\!+\!\delta$ (LP relax.) & midpoint convexity (min) & finiteness & LP duality \\
A4 crush & capacity $\to0$; requirement $\to R$ & infeasible & declared infeasible & certified transfer \\
A5 prohibitive limit & cost or rate of entity $e\to M_1,M_2$ & $\vfun\to\vfun(x_e{=}0)$, two-scale & $x_e{=}0$ feasible & polyhedral threshold \\
A6 annihilation & all requirements $\to0$ & $\vfun=0$ & pure-cost min & direct \\
A7 exchange & swap slot bundles of $e_1,e_2$ & $\vfun$ invariant & congruence & relabeling symmetry \\
\midrule
B1 homogeneity & objective $\times\lambda$ & $\vfun\mapsto\lambda \vfun$ & --- & any MILP \\
B2 relaxation & LP relaxation & $\vfun_{\mathrm{LP}}$ weakly better & MILP & weak duality \\
B3 removal & delete one row & $\vfun$ weakly improves & --- & set inclusion \\
B4 dual gradient & rhs $\pm\delta$ (LP relax.) & subgradient inequality & LP optimal & convexity of $\vfun(b)$ \\
\bottomrule
\end{tabular}}
{}
\end{table}

\subsection{Directional and curvature tests}

\begin{proposition}[Monotone comparative statics]\label{prop:direction}
Let $\Pib$ be role-faithful to $\cA$.  Then, for every slot $j$ and in the
asserted sense $\sigma^\ast$:
(i) if $\texttt{qtype}(j)=\textsf{capacity}$, $\vfun$ is weakly improving in
$\theta_j$; if \textsf{requirement}, weakly worsening;
(ii) if $\texttt{qtype}(j)=\textsf{cost}$ (min) or \textsf{reward} (max),
$\vfun$ is weakly worsening, respectively improving, in $\theta_j$;
(iii) if $\texttt{qtype}(j)=\textsf{rate}$ in a row of sense $\le$
($\ge$), $\vfun$ is weakly worsening (improving) in $\theta_j$;
(iv) if $\texttt{qtype}(j)=\textsf{ratio}$ in a share row of sense $\ge$
($\le$), $\vfun$ is weakly worsening (improving) in $\theta_j$ on $[0,1]$.
All statements hold for mixed-integer candidates.
\end{proposition}

\begin{proposition}[Curvature]\label{prop:curvature}
For any LP (in particular the LP relaxation of a role-faithful candidate),
$b\mapsto \vfun_{\mathrm{LP}}(b)$ is convex if $\sigma^\ast=\min$ and concave
if $\sigma^\ast=\max$, on its effective domain; hence the midpoint predicate
of test A3 holds along every slot segment on which the three probed values
are finite.
\end{proposition}

A3 is run on the relaxation by design.  MILP value functions are not
convex in $b$; the relaxation keeps the test sound while still probing
the candidate's constraint data.

\subsection{Certified crush tests}

Presence testing in \citet{lian2026reloop} perturbs a capacity by
$\times0.001$ and flags the model if the objective moves by less than 5\%.
Soundness fails structurally, because a correct model whose capacity is not
binding, or is protected by a tighter sibling constraint, moves by 0\% and
is flagged.  Our crush test replaces the threshold with a
\emph{certificate} and fires only when the assertion structure itself
guarantees the outcome.

\begin{proposition}[Certified crush]\label{prop:crush}
In annotated mode, suppose the declared structure $S^\ast$ is infeasible
when capacity slot $j$ is set to $0$, or when requirement slot $j$ is set
to a large probe value $R$.  Then every structure-faithful candidate is infeasible at the same
probe.  The certificate is computed by one solver call on $S^\ast$ and the
test by one solver call on the candidate; no thresholds are involved.
\end{proposition}

A candidate that omitted the row containing slot $j$, or bound the slot to
the wrong row, remains feasible at the crush probe and is flagged with
certainty (Theorem~\ref{thm:detect}(ii)).

\subsection{Prohibitive-limit tests}

The subtlest tests exploit an exactness phenomenon.  Beyond a finite
threshold, pricing an activity out of the model is \emph{identical} to
deleting it.

\begin{proposition}[Prohibitive limits, finite threshold]\label{prop:limit}
Let $\Pib$ be role-faithful, let $e$ be an entity whose variables are
$x_e$, and suppose $x_e=0$ is feasible at $\theta^0$
(precondition certified as in Section~\ref{sec:framework}); for
mixed-integer candidates assume the integer variables are bounded, as
they are in every corpus we use.  Write
$\vfun_{e0}$ for the optimal value of the candidate with $x_e$ fixed to $0$.
(i) (Cost side, $\sigma^\ast=\min$.)  If slot set $J_e$ carries $e$'s
objective costs, then there exists $\bar M<\infty$ such that for all
$M\ge\bar M$, replacing $\theta_j\mapsto M$ for $j\in J_e$ yields
$\vfun = \vfun_{e0}$ exactly.
(ii) (Rate side, either sense.)  If slot $j$ is a positive rate of $e$ in a
$\le$-row with nonnegative data, then $\vfun(\theta_j{\to}M)\to \vfun_{e0}$,
with exact equality beyond a finite $\bar M$; for integer $x_e$ the
threshold is explicit, since every $M>b_r$ forces $x_e<1$ and hence
$x_e=0$.
Moreover, for minimization and every $M$ no smaller than the probed
slots' nominal values, $\vfun(\theta_j{\to}M)$ lies between
$\vfun(\theta^0)$ and $\vfun_{e0}$, a two-sided sandwich used as a
runtime check.
\end{proposition}

Because the residual $\vfun_{e0}-\vfun(M)$ of a faithful candidate is
nonnegative, nonincreasing, and convex in $M$ and vanishes beyond
$\bar M$ (E-Companion~\ref{ec:scale}), the implementation solves at two
scales $M_1\ll M_2$ and accepts iff
$|\vfun(M_2)-\vfun_{e0}|\le\max\{\varepsilon,\,4|\vfun(M_1)-\vfun(M_2)|\}$,
a convergence criterion that requires no data-dependent tuning.  Test A6,
which zeroes all requirements and demands the exact value $0$ under a
certified pure-cost precondition, is proved similarly and directly
falsifies wrong-sense and spurious-constant errors.

\begin{remark}[Scale robustness of the two-scale criterion]\label{rem:scale}
The exact predicate of Proposition~\ref{prop:limit} refers to the limit
and is threshold-free; the implemented acceptance rule fixes the probe
scales $(M_1,M_2)$ and a contraction factor of $4$.  Concavity and
piecewise linearity of $\vfun$ in the probe scale imply that the rule is
sound for every faithful candidate whose polyhedral threshold satisfies
$\bar M\le 5M_2-4M_1$ (E-Companion~\ref{ec:scale}); data whose cost
ratios exceed the probe range can defeat it.  The condition is
verifiable, and the boundary sits where the theory places it.  On a
constructed two-cost family the rule passes every faithful instance
with $\bar M$ below the bound, rejects every one above it, and enlarging
$(M_1,M_2)$ to cover the data range restores all passes (E-Companion
Table~\ref{ec:tab:scalebound}).  Soundness of A5 as implemented is
therefore relative to a probe-scale assumption, just as
Proposition~\ref{prop:redundant} makes all guarantees orbit-relative.
\end{remark}

\subsection{Exchange tests}

\begin{proposition}[Exchange invariance]\label{prop:exchange}
Suppose the assertion structure certifies that entities $e_1,e_2$ are
\emph{congruent}: their slot bundles carry identical multisets of
(position, qtype) tags, they participate identically (up to the paired
slots) in every row, and their entity-specific bound rows match in sense and
unslotted data.  Then for every faithful candidate, swapping the two
entities' complete slot-value bundles leaves $\vfun$ unchanged.
\end{proposition}

Exchange tests are the battery's only \emph{cross-slot} relational probes,
and Section~\ref{sec:limits} shows they see errors that every single-slot
probe provably misses.

\subsection{Layer B: coherence without assertions}

Layer B tests are theorems for arbitrary MILPs, covering positive
homogeneity of the value in the objective vector
($\vfun(\lambda c)=\lambda\vfun(c)$), relaxation dominance, removal
monotonicity, and the dual subgradient
inequality $\vfun_{\mathrm{LP}}(b')\ge \vfun_{\mathrm{LP}}(b)+y^\top(b'-b)$
for minimization, valid for any optimal dual $y$ even under degeneracy because
optimal duals are subgradients of the convex function
$\vfun_{\mathrm{LP}}(\cdot)$.  Consequently Layer B has \emph{no} power
against faithful-looking formulation errors, whose corrupted models
satisfy the same theorems (Proposition~\ref{prop:reparam} formalizes the
general phenomenon), but it convicts broken solver interfacing, such as
mis-extracted objective values, inconsistent reported solutions,
sign-flipped duals, and silent presolve failures.  In our pipeline Layer B plays the role that execution-based
checking plays in \citet{lian2026reloop}, with inequalities replacing
statuses.

\begin{theorem}[Battery soundness]\label{thm:soundness}
In exact arithmetic:
(i) every Layer-A test with a certified precondition passes on every
structure-faithful candidate (annotated mode);
(ii) tests A0--A3, A5--A7 pass on every role-faithful candidate whose
preconditions are certified on its own structure (deployment mode);
(iii) Layer-B tests pass on every MILP.
Hence the battery's false-positive rate against the corresponding class is
zero, and every violation constitutes a machine-checkable certificate of
unfaithfulness carrying the offending slot, probe, and property.
\end{theorem}

\section{Detection Limits}\label{sec:limits}

Soundness constrains what a battery may flag; this section characterizes what
it \emph{can} flag.  Fix the probe orbit
$\Theta_{\cB}=\{\tau(\theta^0):\tau\in\cT_{\cB}\}$, the set of instances the
battery may visit, and call two candidates \emph{battery-indistinguishable}
if they produce identical statuses and values on $\Theta_{\cB}$ (and
identical reported duals/incumbents where Layer B reads them).  Since every
predicate is a function of these observations, indistinguishable candidates
receive identical verdicts.  Three invisibility results follow.

\subsection{What no such verifier can see}

\begin{proposition}[Gauge invisibility]\label{prop:gauge}
Let $\Pib'$ be obtained from a faithful $\Pib$ by multiplying the entire
objective (including any constant) by $\lambda>0$.  Then $\Pib'$ passes every
oracle-free test whose predicate is invariant under positive rescaling of all
observed values, a class that contains every test in this paper and every
sign, ordering, ratio, or homogeneity predicate.  Moreover the
transformation is decision-irrelevant: $\Pib'$ has the same optimal-solution
correspondence as $\Pib$ at every $\theta$.
\end{proposition}

The one Layer-A test that pins absolute scale, A6, requires the value $0$,
which is a fixed point of the gauge; a nonzero spurious constant \emph{is}
detected by A6 where certified.  Gauge invisibility is thus benign, since
what the verifier cannot see cannot change any decision.

\begin{proposition}[Orbit redundancy]\label{prop:redundant}
Let $\Pib'$ add to a faithful $\Pib$ any constraints that are inactive at
every $\theta\in\Theta_{\cB}$ (e.g., dominated rows with slack bounded away
from zero on the orbit).  Then $\Pib'$ is battery-indistinguishable from
$\Pib$.  Consequently, detection of superfluous-yet-inactive structure is
possible only by enlarging the orbit, and any fixed battery's guarantee is
intrinsically \emph{relative to its orbit}.
\end{proposition}

This is the honest limit of behavioral verification, which certifies
behavior only where it looks.  The practical mitigation is orbit design.
Our probes already visit the crush and prohibitive-limit extremes,
where mildly redundant rows typically activate, and rows engineered to
remain dominated even there are harmless on the entire probed region.

\begin{proposition}[Reparametrization invisibility and the reach of exchange]\label{prop:reparam}
Let $\Pib'$ misuse a single slot $j$ linearly, so that every occurrence
of $\theta_j$ is replaced by $\lambda\theta_j$ for a fixed $\lambda>0$
with $\lambda\ne1$, the canonical unit error.  Then:
(i) under candidate-side certification, $\Pib'$ passes every
single-slot probe of the battery, covering direction, curvature,
prohibitive limits, and annihilation, and it passes crush probes run
with candidate-side certificates as well, where the requirement-side
case requires $\lambda\ge1$ and the capacity probe value $0$ is a
fixed point.  The reason is that each such predicate is invariant
under an increasing linear reparametrization of the probed axis and every
certificate is evaluated on the reparametrized model itself;
(ii) $\Pib'$ is detected by an exchange test A7 whenever slot $j$ belongs to
a certified-congruent bundle and the swapped instance separates the values,
which holds for generic data;
(iii) if the misuse applies uniformly to \emph{all} objective slots
($\lambda c$), it is the gauge of Proposition~\ref{prop:gauge} and is
invisible to the entire battery.
\end{proposition}

\begin{remark}[Certificates buy power in annotated mode]\label{rem:cert}
In annotated mode the preconditions of the limit and crush tests are
certified on the \emph{declared} structure at $\theta^0$.  A misused
candidate behaves like the faithful family at the point whose $j$th
coordinate is rescaled by $\lambda$, where a certificate valid at
$\theta^0$, say feasibility of $x_e=0$, may fail; the test then fires on a genuine
unfaithfulness that deployment-mode certification would skip.  Declared
assertions thus strictly enlarge the detectable set.  Section~\ref{sec:e1}
quantifies the effect, where certificate transfer convicts an additional
\numMSevenAFiveResiduepct{} of single-slot misuse mutants through the
prohibitive-limit test.
\end{remark}

Proposition~\ref{prop:reparam} yields the sharpest falsifiable prediction
of the theory.  Single-coefficient scale errors must be caught \emph{only}
by A7; uniform objective scaling and orbit-redundant additions,
\emph{never}.
The experiments reproduce this pattern (Table~\ref{tab:matrix}).

\begin{proposition}[Min/max asymmetry of annihilation]\label{prop:asymmetry}
On the cost side of a minimization problem, the prohibitive limit of
Proposition~\ref{prop:limit}(i) is an equality.  Its mirror on the reward
side of a maximization problem, which sets entity $e$'s rewards to zero,
satisfies
only the one-sided bound $\vfun(\text{rewards}_e{\to}0)\ \ge\ \vfun_{e0}$,
which holds for \emph{arbitrary} candidates and hence carries no detection
power; equality can fail for faithful candidates because a zero-reward
activity may still be used to satisfy joint requirements; the
E-Companion gives an example.  Rate-side limits, by
Proposition~\ref{prop:limit}(ii), restore symmetric power for
maximization problems.
\end{proposition}

\subsection{What the battery provably catches}

We now fix the canonical error classes used throughout the empirical
literature and our mutation study
(cf.\ the taxonomy of \citealt{li2026optargus}); $\Pib'$ denotes a faithful
candidate corrupted by one operator.

\begin{theorem}[Class-by-error detectability]\label{thm:detect}
Let the assertion structure certify the stated precondition.  Then the
corrupted candidate is flagged with certainty (in exact arithmetic) as
follows.
\emph{(i) Sense flip} ($\min\leftrightarrow\max$): by A0; by any A2 rate
or ratio probe at which the corrupted response is strict, since
constraint-side monotonicity is inherited from feasible-set inclusion
and therefore reverses with the sense (objective probes carry no power
here: for $x\ge0$ the value is weakly increasing in an objective
coefficient under either sense); and by A6 when certified with
$\vfun>0$.
\emph{(ii) Constraint omission}: by A4 whenever the omitted row's crush
certificate holds, because the corrupted model's value is constant in the
crushed slot and the model therefore remains feasible where infeasibility is
certified.
\emph{(iii) Direction flip} of a slotted row: by A1 at any probe where the
row binds strictly on one side (the response reverses orientation); by A4
when the row carries a crush certificate ($\le$ made $\ge$ is feasible at
zero capacity); by A6 when certified.
\emph{(iv) RHS misassignment} between two slotted rows: by A1 whenever the
two slots' asserted directions differ; by A4 whenever exactly one of them
carries a crush certificate that the swap breaks.
\emph{(v) Dropped objective term} of entity $e$: by A5(i) whenever
$x_e=0$ is certified feasible and using $e$ is strictly profitable in the
corrupted model at the prohibitive probe, i.e.\ $\vfun'_{\lim}<\vfun'_{e0}$.
\emph{(vi) Entity misbinding} (rates of $e_1,e_2$ swapped in one row): by
A5(ii) whenever the certified prohibitive-rate probe of $e_1$ leaves
$x_{e_1}$ strictly active in the corrupted model; by A7 under congruence with
generic data.
\emph{(vii) Broken data binding} (a slot's value hard-coded): by A4/A6 for
rhs slots (the crushed model retains its hard-coded value, contradicting the
certificate) and by A5(ii) for rate slots.  A1 and A2 are silent
because weak monotonicity tolerates constancy; presence must be
established by limit probes.
\end{theorem}

The proofs in E-Companion~\ref{ec:proofs} are short case analyses on top
of Propositions~\ref{prop:direction}--\ref{prop:exchange}; their value is
the \emph{map} they induce from error classes to test classes, which
turns the battery's verdict into a diagnosis, the mutation study into a
controlled test of the theory, and each cell of Table~\ref{tab:matrix}
into a prediction.

\subsection{Thresholds cannot be sound}

\begin{proposition}[No sound nontrivial threshold tester]\label{prop:threshold}
Fix any relative threshold $\tau>0$ and consider the presence tester that
flags a candidate when an extreme perturbation of a capacity slot changes the
optimal value by a factor less than $\tau$
(the CPT rule of \citealt{lian2026reloop}).  For every $\tau>0$ there is a
family of \emph{faithful} two-variable candidates on which the tester's
false-positive rate is one (a capacity that remains slack under the extreme
probe because a sibling bound is tighter); for $\tau=0$ the tester never
flags anything.  Hence no threshold choice yields a sound and nontrivial
presence tester, whereas the certified crush test A4 flags none of these
faithful candidates and retains full power on the corresponding omissions by
Theorem~\ref{thm:detect}(ii).
\end{proposition}

The construction, given in the E-Companion, captures the ordinary
situation in which one resource is abundant.  Empirically, the
reimplemented threshold tester flags \numReloopFPRpct{} of faithful
NL4OPT models for this reason (Section~\ref{sec:experiments}).
\citet{lian2026reloop} recognize the ambiguity of intermediate responses
and route the band between their two thresholds to a non-blocking
informational zone, but responses below the lower threshold still trigger
repair.  The theoretical point is that the ambiguity is an information
problem, which certification resolves and tuning cannot.

\subsection{Cost of verification}

\begin{proposition}[Probe complexity]\label{prop:complexity}
For a candidate with $p$ slots, $E$ entities, $C$ certified congruent
pairs, and $R$ rows, the battery issues $O(p+E+C+R)$ solver calls: $5$
per rhs slot ($2$ incremental solves for A1, whose nominal anchor
$\vfun(\theta^0)$ is computed once per battery run and shared across
slots, and $3$ for A3, whose probe points and relaxation modes differ),
$1$ per coefficient slot (A2), $1$--$2$ per
certified crush row (A4), $3$ per entity and side (A5, two scales plus
the fixed problem), $1$ for A6, $2$ per congruent pair (A7), and $O(R)$
for Layer B.  Because the implementation tests congruence for every
entity pair, $C\le\binom{E}{2}$ and the worst case is $O(p+E^2+R)$; on
the benchmark corpora congruent pairs are rare and the observed cost sits
in the linear regime.  Each call is an LP/MILP of the candidate's own
size.
\end{proposition}

On the benchmark corpus this is \numAvgSolves{} solves and
\numAvgBatterySec{} seconds per model on one CPU core
(Section~\ref{sec:experiments}), three orders of magnitude below the cost of
generating the candidate with a 7B LLM on an A100 GPU.

\section{A Verifiable Modeling Interface}\label{sec:vmi}

Behavioral testing of code silently presupposes a solved problem, because
perturbing the capacity requires knowing which number in the program is the
capacity.  ReLoop extracts this map with an LLM at test time; we make
data binding a first-class, checkable artifact of generation.

\begin{figure}[t]
\FIGURE
{\includegraphics[width=.99\textwidth]{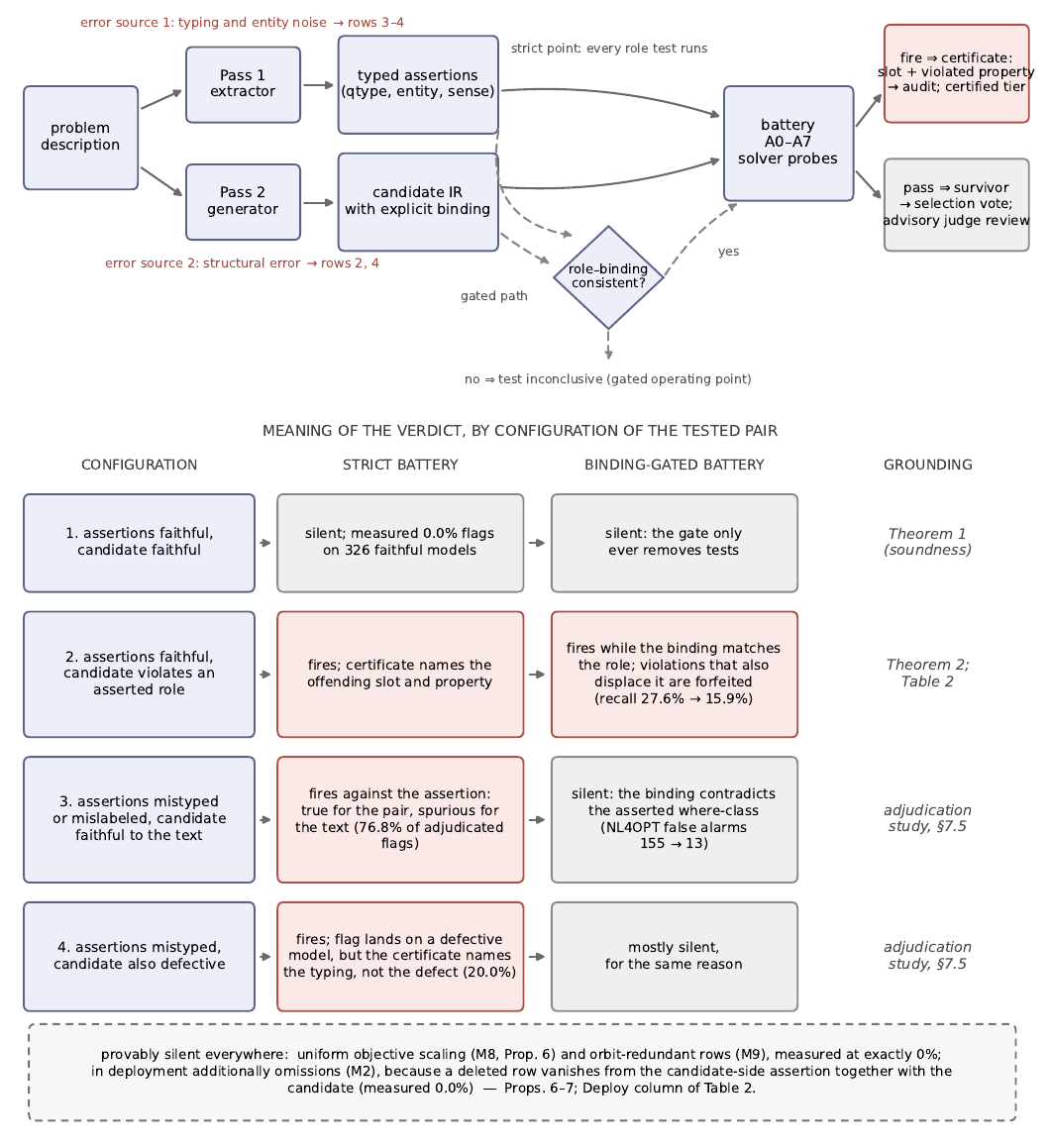}}
{The verifier as a system (top) and the semantics of its verdicts
(bottom).  Solid edges are the strict operating point, dashed edges the
binding-gated one; the two passes create two independent error sources,
which map onto the numbered configurations of the lower tier.
Adjudication shares in rows 3--4 refer to all \numAdjN{}
value-correct-but-flagged NL4OPT certificates, of which a further
\numAdjTRVpct{} expose true role violations; recall figures and flag
counts refer to the NL4OPT deployment run.\label{fig:semantics}}
{The gate never adds a flag, so the zero false-positive guarantee of row 1
is shared by both operating points.}
\end{figure}

\emph{Pass 1 (extraction).}  An LLM reads the description and emits the typed
slot table $\{(s_k,\theta^0_k,\texttt{qtype}_k,e_k)\}$ and the sense
$\sigma^\ast$.  The pass extracts assertions and performs no modeling.
The table is validated against the text (every slot value must appear;
percentages are normalized) and is the sole carrier of semantic
authority thereafter.

\emph{Pass 2 (generation).}  The generator emits the model in a JSON
intermediate representation whose coefficient and right-hand-side fields may
reference quantities \emph{only as slot identifiers}; numeric literals
are reserved for structural constants ($0,\pm1$, ratio complements).
Ratio slots are therefore encoded structurally, never referenced by
id; they are excluded from slot-coverage accounting and are untested by
slot probes in deployment mode.  The IR
compiles deterministically to the solver; slot references resolve to the
extractor's values, so the binding is correct by construction, and models
whose IR references unknown slots, omits declared quantities, or fails
schema checks are rejected before any solve.

The battery then runs in deployment mode (Theorem~\ref{thm:soundness}(ii)):
directional, curvature, limit, annihilation, and exchange tests with
candidate-side certification; crush tests, whose certificates require a
declared structure, are reserved for annotated corpora (the
implementation skips A4 in deployment mode, where certificate and test
would coincide in a single solve).  Beyond enabling
verification, the interface removes a documented failure source,
indexing and data-access errors in free-form solver code
\citep{lian2026reloop}, and makes candidate models diffable and auditable
artifacts.  The two-pass design also isolates responsibility.  Extraction
errors corrupt assertions, which are short structured outputs and hence
votable; generation errors corrupt structure, which is what the battery
tests.

\subsection{Binding-consistent deployment battery}\label{sec:gate}

Deployment-mode guarantees are relative to the extracted assertions, and
the assertions carry two kinds of information of very different
reliability: the slot \emph{values}, which are copied from the text and
validated, and the slot \emph{types and entity labels}, which are genuine
LLM judgments.  The certificate adjudication study of
Section~\ref{sec:e4} shows that when the asserted type disagrees with the
candidate's own binding, the disagreement is overwhelmingly an artifact
of the typing and only rarely an error of the model.  This motivates a
refinement that uses the candidate's explicit binding, which the
interface makes observable, as a consistency gate.  A slot is subjected
to the directional and curvature tests of its asserted role only if the
candidate binds it at least once in the where-class that role implies
(right-hand side for capacities and requirements, coefficients for rates,
objective for costs and rewards).  Otherwise the role premise is already
contradicted by the binding itself and the test is recorded as
inconclusive.  For the exchange class, objective slots
are paired by the entity of the variable they multiply, as constraint
slots are, in place of the extractor's free-text entity
labels.  The gate only ever removes tests, so
Theorem~\ref{thm:soundness} is unaffected and the false-positive rate on
faithful models remains zero by construction; what is traded away is
sensitivity to genuine misuse that happens to coincide with a typing
mismatch.  The strict battery remains the audit instrument of record,
since a strict flag is a true statement about the assertion--candidate
pair even when the assertion is at fault; the gated battery is the
operating point for selection and for triage under noisy extraction.
Figure~\ref{fig:semantics} lays out the complete semantics: which of
the three failure configurations of the assertion--candidate pair fires
which battery, what each flag then means, and where both batteries are
provably silent.  Section~\ref{sec:e4} quantifies both operating
points.

\section{Computational Study}\label{sec:experiments}

The experiments test the theory's predictions, compare the battery
against the natural alternatives, and measure end-to-end value in a
local-model pipeline.  Generation and the LLM baselines run on one A100
GPU, all solves run on CPU cores with HiGHS, and every number in this
section is reproducible from a named artifact file in the replication
package of code, seeds, and per-run outputs that accompanies the paper.

\subsection{Corpora and instrumentation}

\emph{Annotated corpus.}  The NL4OPT generation-track annotations provide,
for each word problem, a declared ground-truth formulation with typed
declarations.  Our parser compiles \numParsedTotal{} of \numAnnTotal{}
dev+test problems (\numParseCovpct{} coverage; failures are tabulated in the
E-Companion) into the IR with slots and assertions derived mechanically from
the declarations.  Of these, \numSeedsTotal{} solve to a finite optimum;
they form the \emph{faithful seeds}.  Whether the battery flags any of
them is the clean false-positive measurement reported below, taken before
any battery-based filtering.

\emph{Synthetic families.}  Controlled production-planning (max) and covering
(min) generators with fully certified preconditions isolate power from
precondition coverage and support size scaling to $40$ variables and $25$
resource rows, integer and continuous.

\emph{Benchmarks with two generators.}  Questions are answered by the
two-pass pipeline of Section~\ref{sec:vmi} with $N{=}8$ candidates per
problem, one greedy and seven sampled at temperature $0.8$, generated
independently by a locally hosted Qwen2.5-7B-Instruct and by
DeepSeek-V4-flash through its public API with identical prompts.  Generation
covers the complete official files of all four benchmarks; the
evaluated pools are the problems with labels and parseable
extractions, namely NL4OPT test (\numBenchNl{} of 289), MAMO EasyLP
(\numBenchME{} of 652), MAMO ComplexLP (\numBenchMC{} of 211), and
IndustryOR (\numBenchIor{} of 100) for the Qwen generator.  The shortfall
from the file sizes is extraction-stage attrition, the pass-1 JSON
failing to parse; it is reported per pool in E-Companion
Table~\ref{ec:tab:yield}, and one pathological MILP instance is excluded
by a solver time-limit rule declared in the replication scripts.  All
selection results below score \emph{all} problems in these pools,
counting a problem with no valid candidate as a miss for every policy.

\emph{Baselines.}  \emph{Threshold behavioral testing} is our
reimplementation of ReLoop's CPT and OPT rules, with $\times0.001$ and
$\times100$ extremes and $\tau_\ell=5\%$, applied to the same IR and slot
table; the port is strictly charitable, since it inherits our exact
binding, and it evaluates ReLoop's \emph{detector} in isolation from
its repair loop.  \emph{LLM-as-judge} asks
Qwen2.5-7B-Instruct, DeepSeek-R1-Distill-Qwen-7B, and the
frontier-class DeepSeek-V4-flash to verdict each candidate, rendered as
question plus canonical LP, as CORRECT or INCORRECT.  \emph{Execution
consensus} is majority voting over candidate optimal values, the standard
selection heuristic.

\input{tables_gen}

\subsection{E1: the detectability matrix}\label{sec:e1}

For each faithful seed and each of eleven mutation operators
(Table~\ref{tab:matrix}), up to three distinct single-error mutants are
generated, \numMutantsTotal{} in all, and passed to the battery in
annotated mode.  The operators map onto Theorem~\ref{thm:detect} and
Propositions~\ref{prop:gauge}--\ref{prop:reparam}: five detectable classes
(M1 direction flip, M2 omission, M3 rhs swap, M4 dropped objective term, M5
sense flip), two binding breaks (M10 hard-coded rhs, M11 hard-coded
coefficient), one conditional class (M6 entity misbinding), and three
theory-invisible controls (M7 single-slot scale misuse, exchange-only by
Proposition~\ref{prop:reparam}; M8 uniform objective scaling; M9
orbit-redundant row).

Throughout the empirical sections we fix three mutant populations.  The
\emph{core} population comprises classes M1--M5 ($n=\numCoreN$); the
\emph{conditional} population comprises the mutants of M2, M4, M6, M10,
and M11 whose stated preconditions certify on the specific mutant
($n=\numCondN$, against $n=\numUncondN$ uncertified); the
\emph{blind-analysis} population comprises all eight detectable classes
M1--M6, M10, and M11 ($n=\numBlindSetN$) and supports the
execution-blindness statistics.  Percentages refer to the named
population.

Three findings stand out (Table~\ref{tab:matrix}).  First,
\emph{soundness holds in practice}.
The battery flags \numCleanFPRpct{} of the \numSeedsTotal{} faithful
seeds, while the threshold tester flags \numReloopFPRpct{} of the same
models.  Second, \emph{the detectability matrix matches the theory}.
M8 and M9 sit at exactly $0\%$ across all test classes; M7 is detected in
\numMSevenDetpct{} of mutants, all but a \numMSevenAFiveResiduepct{}
residue through the exchange class A7 as Proposition~\ref{prop:reparam}
requires, the residue itself explained and predicted by the
certificate-transfer mechanism of Remark~\ref{rem:cert}; M5 is caught at
$100\%$.  Certification governs the conditional classes asymmetrically.
Certified omissions (M2) are caught at \numMTwoCondDetpct{}
($n=\numMTwoCondN$), matching the certainty clause of
Theorem~\ref{thm:detect}(ii), and uncertified entity misbinding (M6) at
exactly \numMSixUncondDetpct{}, matching the necessity of its
precondition; between these poles, certified detection is
\numMTenCondDetpct{} for M10, \numMElevenCondDetpct{} for M11,
\numMFourCondDetpct{} for M4, and \numMSixCondDetpct{} for M6, short of
certainty because Theorem~\ref{thm:detect} attaches additional
strictness and genericity conditions to these classes that the
certificates do not check.  Pooled, certification raises detection from
\numUncondDetpct{} to \numCondDetpct{}.  Third, \emph{execution accuracy
is blind where the battery is not} (Figure~\ref{fig:blind}).
\numBlindSharepct{} of the blind-analysis population, \numBlindN{}
mutants, reproduce the correct optimal value at the nominal instance, so
label comparison cannot separate them from correct models even with
perfect ground truth, yet the battery convicts \numBlindDetpct{} of them.
This population is the sharpest evidence that verification and
evaluation are different problems.

A fourth finding quantifies the \emph{deployment gap}.  Rerunning the
identical mutant set with the verifier restricted to the information
available in deployment, candidate-side assertions only, A4 skipped, and
ratio slots untestable, lowers pooled detection from \numAnnotAllpct{} to
\numDeployAllpct{} (Deploy column of Table~\ref{tab:matrix}) while the
clean false-positive rate remains \numDeployCleanFPRpct{}.  The gap
follows the information structure.  Omissions (M2) drop from
\numMTwoCondDetpct{} when certified to exactly \numDeployMTwopct{},
since a deleted row disappears from the candidate-side assertion
together with the candidate and the two are self-consistent, which is
why crush certification requires a declared structure.  Direction
flips, rhs swaps, dropped objective terms, and hard-coded rhs roughly
halve.  Sense flips stay at $100\%$ because the extractor's sense
survives as an independent assertion, and the entity-misbinding,
hard-coded-coefficient, and invisibility rows are unchanged.  The
annotated-mode matrix is therefore an upper envelope
whose deployment projection is itself predictable, and the two columns
bound what a practitioner should expect with and without a declared
reference structure.

\subsection{E2: baselines on equal footing}\label{sec:e2}

The panel comprises \numPanelMutants{} mutants stratified over all eleven
operators (twelve per operator, including the invisibility controls) and
\numPanelClean{} faithful models drawn from the battery-passing seeds.
Because the panel cleans are battery-passing seeds, their measured flag
rate restates E1's clean false-positive result; Figure~\ref{fig:fpr}
therefore labels each bar with its own denominator.  On the full panel the battery attains
recall \numOursDetpct{} at \numOursFPpct{} false positives.  The
threshold tester and the Qwen judge flag more mutants (recall
\numReloopDetpct{} and \numJudgeDetpct{}) at false-positive rates that
disqualify them as auditors (\numReloopPanelFPpct{} and
\numJudgeFPpct{}); the R1-distill judge is conservative (recall
\numJudgeTwoDetpct{} at \numJudgeTwoFPpct{} false positives) but offers
neither certificates nor slot-level localization.  Because the panel is
mutant-heavy, prevalence-dependent scores such as $F_1$ reward
indiscriminate flagging (Qwen judge \numJudgeF{}, threshold
\numReloopF{}, battery \numOursF{}, R1 judge \numJudgeTwoF{}); by the
prevalence-free Youden index $J=\text{recall}-\text{FPR}$ the battery
ranks first among the four systems compared so far (\numOursJ{} versus
\numReloopJ{}, \numJudgeJ{}, \numJudgeTwoJ{}).  The panel labels every
mutant as faulty, yet M8 is provably decision-irrelevant
(Proposition~\ref{prop:gauge}) and M9 is orbit-invisible, so treating
them as faults penalizes a sound verifier for correctly staying silent.
Excluding the twenty-four M8/M9 control mutants
($n=\numPanelMutantsNC{}$) raises the battery to recall
\numOursDetNCpct{} ($J=\numOursJNC$) and the threshold tester to
\numReloopDetNCpct{} ($J=\numReloopJNC$); the LLM judges are shown on
the full panel.  A frontier-class judge raises the judge ceiling
substantially.
Single-sample DeepSeek-V4-flash reaches recall \numJudgeVThreeDetpct{}
at \numJudgeVThreeFPpct{} false positives ($J=\numJudgeVThreeJ$;
\numJudgeVThreeDetNCpct{} and $J=\numJudgeVThreeJNC$ without the
controls); a five-sample self-consistency vote sharpens this to recall
\numJudgeSCDetpct{} at \numJudgeSCFPpct{} false positives
($J=\numJudgeSCJNC$ without the controls), the best Youden index on the
panel, while a long-reasoning variant at a matched prompt trades a
higher false-positive rate (\numJudgeRznFPpct) for no recall gain over
the vote (per-item verdicts in the artifacts).  The accusations these
judges get wrong are instructive.  Most condemn faithful continuous
models for lacking integrality that the ground-truth annotations
themselves do not impose, and they deliver these verdicts with the same
confidence as their correct ones.  Certified flags and judge opinions
are complementary evidence and compose into a triage, certified
convictions first, advisory review by the self-consistency judge
second; the composition convicts \numEscRecallNCpct{} of detectable
panel mutants (\numJBBoth{} by both, \numJBJudgeOnly{} by the judge
alone, \numJBBatOnly{} by the battery alone) and confines every false
accusation to the advisory tier.
Only the battery's flags carry certificates.  A battery flag proves
unfaithfulness relative to the assertions and names the offending slot
and property, while judge flags are opinions and threshold flags are
heuristics, and the distinction is operationally decisive when flags
trigger costly repair or human escalation
\citep{huang2024llmcannotselfcorrect}.

\subsection{E3: power, budget, and scale}

The synthetic family separates power from precondition coverage.  None
of the \numSynthSolvable{} solvable synthetic models is flagged.
On the eight detectable operators the full battery convicts
\numSynthDetFullpct{} of mutants ($n=\numSynthDetFullN$); under a
solver-call budget, detection reaches $95\%$ of that full-battery power
by \numBudgetNinetyFive{} solves and gains little thereafter, against a
full-battery cost of roughly $105$ solves at this size
(Figure~\ref{fig:budget}, which also shows the all-operator curve
including the invisibility controls).  Battery cost grows near-linearly
in model size with millisecond-scale LPs up to the largest sizes tested.
The two-scale limit criterion is validated across four decades of $M$
with one fixed misbinding mutant per seed reused at every scale pair.
All \numFiniteMFaithN{} faithful models pass at every pairing, and the
\emph{same} \numFiniteMDetpct{} of the \numFiniteMMutN{} mutants is
flagged at every pairing, so the criterion's verdict is scale-stable on
this family (E-Companion Table~\ref{ec:tab:finiteM}); its scale limits
are mapped by Remark~\ref{rem:scale} and E-Companion
Table~\ref{ec:tab:scalebound}.

\subsection{E4: verification as selection, adjudication, and audit}\label{sec:e4}

For each benchmark family, the deployment-mode battery scores the $N=8$
candidates.  We evaluate hard filtering, a majority vote among battery
survivors, in both the strict and the binding-gated variants of
Section~\ref{sec:gate}; a \emph{weighted} vote, with each candidate's
vote damped by $2^{-\#\text{violations}}$; the aggressive threshold
prefilter; and a \emph{cascade} that applies the threshold prefilter and
then the violation-weighted vote.  The comparators are greedy selection,
uniform-random selection among valid candidates, and plain majority
voting, with the oracle-any hit rate as headroom; filters fall back to
plain majority when no candidate survives (Table~\ref{tab:select}).
A candidate is scored value-correct when its
optimum matches the numeric label at relative tolerance $10^{-4}$
(labels serialized as digit strings are parsed as numbers); the label
``No Best Solution'' requires the candidate to be infeasible or
unbounded.  We report the all-problems denominator, on which a problem
with no valid candidate scores zero for every policy including the
oracle; E-Companion Table~\ref{ec:tab:yield} reports pipeline yields.

\emph{Selection.}  Soundness costs nothing as a selector, and where the
gate applies it starts to pay.  On the pools where consensus is
informative the weighted vote is statistically indistinguishable from
plain majority (within $1.2$ points; McNemar $p>0.15$ on every pool),
the strict hard filter gives up about
three points on NL4OPT (\numSelBatNlpct{} against \numSelMajNlpct),
and the gated filter edges past majority at \numSelGateNlpct{}
(\numGateSelWin{} paired wins against \numGateSelLose{}, McNemar
$p=\numGateSelP$).  The
threshold prefilter is significantly destructive where consensus
works.  On MAMO EasyLP it scores \numSelReloopMEpct{} against majority's
\numSelMajMEpct{}, losing \numReloopMELose{} paired problems while
winning \numReloopMEWin{} ($p<10^{-4}$), because a
\numReloopFPRpct{}-false-positive filter deletes correct candidates from
the vote.  On MAMO ComplexLP and IndustryOR every policy compresses
against a low oracle ceiling (\numSelOracleMCpct{} and
\numSelOracleIorpct{}) and no pairwise difference is significant.
Soundness and selection utility remain different objectives; what
changes with the gate is that they stop being in tension.

\emph{Adjudicating every value-correct flag.}  The strict battery flags
\numVCFlaggedNl{} of the \numVCCorrectNl{} value-correct NL4OPT
candidates (\numVCFlagShareNl{}; \numVCFlagSharePooled{} pooled over the
four families, \numVCFlaggedPooled{} of \numVCCorrectPooled{}), each
flag a certificate of role violation relative to the extracted
assertions.  We adjudicated all \numAdjN{} NL4OPT cases individually
against the problem text, under a two-stage protocol: an LLM first pass
over every case, then human audit of every true-violation and
low-confidence case and a random sample of the remainder (per-case
records in the artifacts; E-Companion Table~\ref{ec:tab:adjudication}).
The adjudication yields three findings.  First, \numAdjTRV{} certificates
(\numAdjTRVpct) expose genuine role violations that the execution
metric rewards: a capacity written into a $\ge$ row that happens to be
active at equality, requirements bound to equality rows that turn
infeasible when the requirement is relaxed, capacity values wired in as
coefficients.  Second, a further \numAdjExtBadpct{} of the flagged
candidates are independently unfaithful to the text, with invented,
vacuous, or wrong-direction rows whose optimal value coincides by
accident, so the flag lands on a defective model even though the
certificate names a different defect than the text reveals.  Third, the
remaining \numAdjExtOKpct{} trace to three identifiable extraction-noise
mechanisms: per-unit rates typed as requirements or capacities, which
fire A1/A3 through coefficient bindings; objective-typed cost slots that
the text places in budget rows, which fire A2; and entity labels
finer-grained than the variables they describe, which break A7 pairing.
All three mechanisms contradict the candidate's own binding, which is
the condition the gate checks
(Figure~\ref{fig:semantics}): gating cuts the NL4OPT
false-alarm count from \numVCFlaggedNl{} to \numGateFPNl{} and lifts
flag precision from \numFlagPrec{} to \numGatePrecNlpct{}, at a recall
cost of \numFlagRec{} to \numGateRecNlpct{}.

\emph{What a flag costs and buys.}  A screening model prices the trade.
Let reviewing one flag cost one unit and shipping one wrong model cost
$\rho$ units.  Pooled over the four Qwen pools, battery screening beats
no screening once $\rho>\numRhoStarNone$, about one and a half reviews
per avoided miss.  The threshold detector raises recall but issues
\numReloopFlagMult$\times$ the flags (\numReloopFlagsPooled{} against
\numBatFlagsPooled) with over four times the false alarms, and its
expected cost drops below the battery's only when misses cost at least
$\numRhoStarReloop\times$ a review, before pricing the fact that a
certificate localizes the repair while a threshold flag does not.

\emph{A stronger generator.}  With DeepSeek-V4-flash as the generator the
oracle ceiling rises from \numSelOracleNlpct{} to \numDSOracleNlpct{} on
NL4OPT and from \numSelOracleMEpct{} to \numDSOracleMEpct{} on MAMO
EasyLP, and the selection policies compress to within half a point of
one another (E-Companion Table~\ref{ec:tab:deepseek}).  The audit role
sharpens.  On MAMO ComplexLP the battery flags
seventeen DeepSeek candidates, every one of them value-wrong; on NL4OPT,
where nearly every candidate is value-correct, the strict battery's
flags concentrate on extraction artifacts while the gated battery flags
a single candidate, the operating profile a sound auditor should show
when there is little left to find.

\emph{A worked audit beyond the corpora.}
E-Companion~\ref{ec:case} audits end to end a depot-location description
with binary operate decisions, fixed charges, and capacity-linking rows,
a structural family the benchmarks do not contain.  The one faithful
candidate passes both operating points while the threshold tester
false-flags it on the closed depot's slack capacity, the
abundant-resource mechanism of Proposition~\ref{prop:threshold} arising
unprompted; an injected transposition of a capacity with a requirement
is convicted by certificates that name the transposed pair; and
corruptions of the unslotted linking row stay silent, as the Deploy
column of Table~\ref{tab:matrix} predicts.

\emph{Frontier judges in deployment.}  Applied to the same
\numValidCands{} Qwen candidates, the DeepSeek-V4-flash judge is the strongest
selector we measure, \numJudgeDeploySelNlpct{} on NL4OPT against
majority's \numSelMajNlpct{}, and as a detector it reaches
\numJudgeDeployRecpct{} recall at \numJudgeDeployPrecpct{} precision.
Its accusations land on \numJudgeDeployFPRpct{} of all value-correct
candidates, roughly twice the strict battery's pooled share
(\numVCFlagSharePooled) and five times the gated battery's
(\numGateVCFlagSharePooled), each delivered without a certificate.
The comparison sharpens a division of labor.  Judge opinions rank
candidates well and cost a frontier-model call per candidate; battery
certificates prove violations for \numAvgSolves{} millisecond-scale LP
solves; and the two compose into the triage quantified on the panel in
Section~\ref{sec:e2}.  Run as a selector in deployment, the same triage
tracks the judge-alone policy (\numTriageSelNlpct{} against
\numJudgeDeploySelNlpct{} on NL4OPT) while its certified tier retains
the zero-false-positive guarantee that neither the judge nor the
threshold rule can offer.

Because the battery never consults labels, the same machinery doubles as
an \emph{audit}.  Candidates that pass the full battery yet disagree with
the published label localize suspicious labels, and cross-checking the
NL4OPT annotations' own optima against the labels isolates outright
inconsistencies between the two official artifacts.  Eighteen of 207
matched test problems fail a mechanical both-solve criterion, the
annotation's LP and all-integer optima both differing from the published
label; the cases are catalogued with notes in E-Companion
Tables~\ref{ec:tab:audit} and~\ref{ec:tab:auditdetail}, consistent with
the label-noise reports of \citet{chen2025solverinformed}.

\subsection{Scope}

The theory is exact for the MILP families the benchmarks occupy and for
the assertion vocabulary of Section~\ref{sec:framework}; nonlinearities
beyond positive homogeneity, uncertainty sets, and multi-objective
structures need their own necessary-condition catalogs, which the
framework accepts as new test classes.  Deployment-mode guarantees are
relative to extracted assertions, which are short, structured, and
votable, and the adjudication study quantifies how typing noise in
extraction accounts for the spurious flags that remain.  The blind set
of Section~\ref{sec:limits} is provably nonempty:
Proposition~\ref{prop:redundant} makes all guarantees orbit-relative,
Remark~\ref{rem:scale} makes the implemented limit tests
probe-scale-relative, and the deployment column of
Table~\ref{tab:matrix} shows that candidate-side certification
structurally forfeits omission detection.  The computational evidence
covers two generators, one local 7B model and one frontier API model,
and four LP/MILP word-problem benchmarks, all through the two-pass
interface, with extraction-stage attrition reported per pool and the
adjudication protocol pairing an LLM first pass with human audit.
Throughout, the unit of evaluation is the detector, whose certificates
hand any downstream repair policy a localized defect.

\section{Conclusion}\label{sec:conclusion}

Fifty years of sensitivity analysis were built to answer how an optimal
plan changes when the data change, for models assumed correct.  This paper
reuses that theory, unchanged, to decide whether a model is right at all,
for generators assumed fallible.  The reuse yields sound tests that
cannot raise a false alarm, provable detection limits, and a
detectability map that the experiments reproduce, including its zeros.
The practical payoff is a verifier with no data-dependent tuning.  It
audits at a false-positive rate that none of the perturbation rules or
LLM judges we measure attains, it composes with those judges as the
certified tier of a triage, and it exposes benchmark label defects as a
by-product.  The conceptual payoff is larger.  Verification of
AI-generated optimization models is, to a useful extent, an
\emph{optimization} problem, and classical theory already covers much
of its tractable core.

\ACKNOWLEDGMENT{The authors thank the maintainers of the NL4OPT, MAMO, and
IndustryOR benchmarks and of the HiGHS solver.  All experiments used
publicly available data and locally hosted open-weight models.}

\bibliographystyle{informs2014}
\bibliography{references}

\newpage
\ECSwitch

\ECHead{E-Companion: Proofs, Constructions, and Additional Results}

\section{Proofs}\label{ec:proofs}

Throughout, $\Pib$ is the candidate \eqref{eq:candidate}, $\vfun$ its value
function with the infinity conventions of Section~\ref{sec:framework}, and
the words improves and worsens are read in the asserted sense
$\sigma^\ast$ (for minimization smaller is better and $+\infty$ is worst;
mirrored for maximization).  We use repeatedly that if
$\cX'\subseteq\cX$ then optimizing over $\cX'$ is weakly worse, for any
objective, including over integer-restricted sets.

\subsection{Proof of Proposition \ref{prop:direction}}
(i) Let slot $j$ be a \textsf{capacity}, entering (role-faithfulness) only as
the RHS of one row $a^\top x\le\theta_j$.  For $\theta_j'\ge\theta_j$ the
constraint set satisfies
$\cX(\theta)\subseteq\cX(\theta')$ pointwise in all other coordinates, since
$a^\top x\le\theta_j\le\theta_j'$; hence $\vfun(\theta')\le\vfun(\theta)$ for
$\min$ (and $\ge$ for $\max$): $\vfun$ weakly improves.  A
\textsf{requirement} slot enters as $a^\top x\ge\theta_j$; raising $\theta_j$
shrinks $\cX$, so $\vfun$ weakly worsens.  Infeasibility and unboundedness
are covered by the conventions since inclusion of feasible sets preserves the
extended-real ordering of optimal values.

(ii) Let slot $j$ be a \textsf{cost} of entity $e$ in a $\min$ problem,
entering only as the objective coefficient of $x_e\ge0$:
$f(x;\theta)=g(x)+\theta_j x_e$.  For each fixed feasible $x$,
$f(x;\cdot)$ is nondecreasing in $\theta_j$ because $x_e\ge0$; the pointwise
infimum of nondecreasing functions is nondecreasing, so $\vfun$ weakly
worsens.  \textsf{reward} in a $\max$ problem is the mirror statement with
suprema.

(iii) Let slot $j$ be a \textsf{rate} in row $r$ of sense $\le$:
$\theta_j x_e+(\text{rest})_r\le b_r$.  If $\theta_j'\ge\theta_j$, any $x\ge0$
feasible at $\theta'$ satisfies
$\theta_j x_e\le\theta_j' x_e\le b_r-(\text{rest})_r$, hence is feasible at
$\theta$: $\cX(\theta')\subseteq\cX(\theta)$ and $\vfun$ weakly worsens.
Sense $\ge$ is the mirror inclusion.

(iv) A \textsf{ratio} slot $\rho\in[0,1]$ of entity $e$ in a $\ge$-share row
reads, canonically, $(1-\rho)x_e-\rho\sum_{w\ne e}x_w\ge0$, i.e.\
$x_e\ge\rho\,\mathbf{1}^\top x$.  For $\rho'\ge\rho$ and $x\ge0$,
$x_e\ge\rho'\mathbf{1}^\top x\ \Rightarrow\ x_e\ge\rho\,\mathbf{1}^\top x$
because $\mathbf{1}^\top x\ge0$; the feasible set shrinks and $\vfun$ weakly
worsens.  Sense $\le$ mirrors.  Integrality never enters any argument above.
\Halmos

\subsection{Proof of Proposition \ref{prop:curvature}}
Write the LP relaxation in the form
$\vfun_{\mathrm{LP}}(b)=\min\{c^\top x: Ax\le b,\ Bx = d,\ x\ge0\}$ (rows of
sense $\ge$ are negated into $\le$; the probed slot moves one coordinate of
$b$ along a segment).  LP duality gives, on the domain where the primal is
feasible and bounded,
$\vfun_{\mathrm{LP}}(b)=\max\{\,b^\top y + d^\top z:\ A^\top y+B^\top z\le c,\
y\le0\,\}$, a pointwise supremum of affine functions of $b$; hence
$\vfun_{\mathrm{LP}}$ is convex on its effective domain, and along any
segment $b(\theta_j)$ affine in $\theta_j$ the univariate restriction is
convex, which is the midpoint predicate.  If $\sigma^\ast=\max$, the same
argument on $-\vfun$ yields concavity.  (The test fires only when all
three probed values are finite.)\Halmos

\subsection{Proof of Proposition \ref{prop:crush}}
By structure-faithfulness, $\cX_{\Pib}(\theta)=\cX_{S^\ast}(\theta)$ for all
probe instances.  The certificate is the solver's proof that
$\cX_{S^\ast}(\theta^0_{-j},0)=\emptyset$ (capacity crush; requirement crush
with value $R$ analogous).  Then $\cX_{\Pib}(\theta^0_{-j},0)=\emptyset$: the
candidate must report infeasible.  The test never fires without the
certificate, so on certified instances it cannot flag a structure-faithful
candidate.\Halmos

\subsection{Proof of Proposition \ref{prop:exchange}}
Under the certified congruence, define the swap $\tau$ that exchanges the
two entities' slot-value bundles and the relabeling $\pi$ that exchanges
their variables.  Congruence states that applying $\tau$ and $\pi$ together
maps the candidate's data $(c,c_0,A,b)$ at $\theta^0$ to itself: paired
objective slots exchange coefficients of exchanged variables, paired rate
slots exchange row entries of exchanged variables within each row, paired
bound rows exchange as whole rows, and all unslotted entries touching the
two entities agree by assumption.  Hence the program at the swapped
instance equals the original program up to the variable relabeling $\pi$.
A relabeling is a bijection between the two feasible sets that preserves
objective values and integrality, so the optimal values coincide, which is
the predicate of test A7.\Halmos

\subsection{Proof of Proposition \ref{prop:limit}}
We give the argument for (i), $\sigma^\ast=\min$; (ii) is analogous with the
constraint squeezing $x_e$ in place of the objective penalty.

\emph{Sandwich.}  Fix $M\ge0$ and let $\vfun(M)$ denote the value at the
probe $\theta_j\mapsto M$ for $j\in J_e$.  Any $x$ feasible with $x_e=0$ has
probe objective equal to its original objective (role-faithfulness: the
probed slots multiply only $x_e$'s coordinates), so
$\vfun(M)\le\vfun_{e0}$.  Conversely the probe only increases objective
coefficients of nonnegative variables (for $M\ge\theta^0_j$), so
$\vfun(M)\ge\vfun(\theta^0)$ by Proposition~\ref{prop:direction}(ii): the
claimed two-sided sandwich.

\emph{Limit and finite threshold.}  Because $x_e=0$ is feasible,
$\vfun(M)\le\vfun_{e0}<+\infty$ for all $M$.  Consider first the LP
relaxation case ($e$ continuous).  The probe objective is
$c_{-e}^\top x_{-e} + M\,\mathbf{c}_e^\top x_e$ with
$\mathbf{c}_e>0$ on $e$'s coordinates (values $\theta_j$ replaced by the
common scale $M$; taking all $j\in J_e$ to a common $M$ is what the
implementation does).  Parametric LP with a scalar parameter multiplying a
fixed objective block admits finitely many optimal bases: there exist
breakpoints $0=M_0<M_1<\dots<M_K<\infty$ such that on
$(M_K,\infty)$ some fixed basis $B^\ast$ is optimal.  If the corresponding
optimal solution had $x_e\ne0$, its objective would grow without bound in
$M$ while $\vfun(M)\le\vfun_{e0}$ stays bounded, a contradiction.  Hence for
$M>M_K=:\bar M$ an optimal solution has $x_e=0$, and its objective is then
independent of $M$ and equal to the best $x_e=0$ objective, i.e.\
$\vfun(M)=\vfun_{e0}$ exactly.  For integer $x_e$ (case (ii) with a $\le$-row
of nonnegative data), feasibility forces
$x_e\le b_r/M$ once $\theta_j=M$; $M>b_r$ forces $x_e<1$, hence $x_e=0$,
and the probed program then coincides with the $x_e=0$ program, which
gives the explicit threshold in the statement.  For mixed problems: the sandwich keeps $\vfun(M)$ within
$[\vfun(\theta^0),\vfun_{e0}]$; on the branch $x_e\ge1$, every feasible
point's objective is at least $M$ times a positive quantity plus a finite
branch infimum (a negative-objective recession direction of the branch
with $x_e$ fixed would be a recession direction of the nominal problem as
well, contradicting finiteness of $\vfun(\theta^0)$), so beyond a finite
threshold every optimum lies in the $x_e=0$ branch, where the
parametric-LP argument applies for each integer assignment; by the
boundedness assumption on the integer variables there are finitely many
assignments, each eventually constant in $M$, so their minimum is
eventually constant as well, with $\bar M$ the largest relevant
breakpoint.

\emph{Rate side for either sense.}  With slot $j$ a positive rate of $e$ in a
$\le$-row of nonnegative data, any feasible $x$ satisfies
$x_e\le b_r/M\to0$; feasibility at $x_e=0$ is preserved, and by continuity of
the (relaxed) value in the box constraint $x_e\le b_r/M$ together with the
same finite-basis argument, $\vfun(M)=\vfun_{e0}$ beyond a finite threshold;
the maximization case uses the identical squeeze (no objective sign enters).
\Halmos

\subsection{Soundness of A6 and proof of Theorem \ref{thm:soundness}}
For A6, the certified preconditions are: $\sigma^\ast=\min$; objective a
nonnegative combination of $x\ge0$ with zero constant; after zeroing all
\textsf{requirement} slots, every $\ge$-row of the \emph{declared} structure
has RHS $\le0$, every $\le$-row RHS $\ge0$, every equality RHS $=0$, and all
lower bounds are $0$.  Then $x=0$ is feasible for any structure-faithful
candidate at the probe and its objective is $0$; nonnegativity of the
objective gives $\vfun=0$ exactly.

Theorem~\ref{thm:soundness}: (i) each Layer-A test was proved sound for
structure-faithful (A4, A6, A7 with declared-side certificates) or already
for role-faithful (A0--A3, A5) candidates; structure-faithfulness plus the
declared origin of assertions implies role-faithfulness for the slotted
positions, so all tests pass.  (ii) In deployment mode the certificates
(entity-zero feasibility for A5; congruence for A7; the structural
preconditions of A6) are computed on the candidate's own structure; for a
role-faithful candidate these coincide with the true properties, and each
test's proof applies verbatim.  (iii) Layer-B statements are classical
theorems for arbitrary MILPs: positive homogeneity of optimal values;
relaxation dominance; removal monotonicity (set inclusion); and the
subgradient inequality, valid for \emph{every} optimal dual $y$ of the
relaxation because $\vfun_{\mathrm{LP}}$ is convex and $y$ supports it at
$b$.  Degeneracy affects uniqueness of $y$, not the support property.
The one-sided implemented tolerances only widen the pass region.
\Halmos

\subsection{Proof of Proposition \ref{prop:gauge}}
Observed statuses are invariant: scaling the objective by $\lambda>0$ changes
neither feasibility nor boundedness.  Every observed finite value scales by
$\lambda$, and $\pm\infty$ are fixed.  Each predicate class is invariant:
orderings and one-sided comparisons ($u\le w+\varepsilon(u,w)$ with
homogeneous tolerance), midpoint inequalities, equalities between observed
values (A5, A7), the value $0$ (A6), statuses (A4), homogeneity (B1,
$\vfun(3\lambda c)=3\vfun(\lambda c)$), dominance and removal
comparisons, and the subgradient inequality (duals scale with the
objective).  Decision irrelevance: $\arg\min$ sets of $f$ and $\lambda f$
coincide on any set.\Halmos

\subsection{Proof of Proposition \ref{prop:redundant}}
By assumption the added rows are inactive at every probe instance: removing
them does not change the feasible set at any $\theta\in\Theta_{\cB}$,
hence statuses, optimal values, optimal solutions, and (choosing the duals
that assign them zero multipliers, as any vertex-optimal basis of the
augmented system may) the observed duals coincide with the faithful model's.
All predicates are functions of these observations.\Halmos

\subsection{Proof of Proposition \ref{prop:reparam}}
Write the misused model as $\Pib'(\theta)=\Pib(\lambda\ast_j\theta)$ where
$\lambda\ast_j$ multiplies coordinate $j$ by $\lambda$ and $\Pib$ is
faithful.

(i) Every single-slot test on slot $j$ observes
$\vfun'(\theta_j{=}t)=\vfun(\lambda t)$ along the $j$-axis (other slots at
nominal), i.e.\ the faithful value function precomposed with an increasing
linear map of the probed coordinate.  Directional predicates compare $t_1<t_2$:
$\lambda t_1<\lambda t_2$ preserves orientation, so the weak inequalities of
Proposition~\ref{prop:direction} hold.  The midpoint test uses equally spaced
$t$-triples, which remain equally spaced (with spacing $\lambda\delta$) after
the map, and convexity is preserved under affine precomposition.  Crush maps
$0\mapsto0$, so the certified status transfers.  Prohibitive limits map
$M\mapsto\lambda M\to\infty$; both the limit value $\vfun_{e0}$ (unaffected:
$x_e=0$ solutions do not see slot $j$ when $j\in J_e$; when $j$ is a rate of
$e$, likewise) and the two-scale residual criterion (residuals
$r(M)=\vfun(\lambda M)-\vfun_{e0}$ still vanish beyond $\bar M/\lambda$ and
decay proportionally) are preserved.  Annihilation (A6) probes requirement
slots at $0$, again a fixed point.  Tests on slots $k\ne j$ observe the
faithful family with the $j$th coordinate frozen at $\lambda\theta^0_j$, which is a
faithful family with different nominal data, so they pass by
Theorem~\ref{thm:soundness}.  (iii) is Proposition~\ref{prop:gauge}.

(ii) The exchange test swaps the bundles of the entity $e_1$ carrying
slot $j$ and its congruent partner $e_2$ carrying slot $j'$.  The misused model evaluates
$\vfun^{swap,\prime}=\vfun(\dots,\lambda\theta^0_{j'}\text{ at }j,\
\theta^0_j\text{ at }j',\dots)$, whereas invariance (which the predicate
asserts, and which holds for the faithful model by
Proposition~\ref{prop:exchange}) requires
$\vfun^{swap,\prime}=\vfun'(\theta^0)=\vfun(\dots,\lambda\theta^0_j,\
\theta^0_{j'},\dots)$.  The two right-hand sides are the faithful value at
two \emph{different} data points whenever
$\lambda\theta^0_{j'}\neq\lambda\theta^0_j$ i.e.\ $\theta^0_j\ne\theta^0_{j'}$;
equality of the values is then a nontrivial equality of a piecewise-linear
concave/convex function at two distinct arguments, which fails for data
outside a finite union of lower-dimensional sets (where the two points share
a linearity piece with gradient orthogonal to their difference).  The word
generic in the statement refers to this measure-zero exclusion.\Halmos

\subsection{Proof of Proposition \ref{prop:asymmetry}}
One-sidedness for arbitrary candidates: any $x$ feasible with $x_e=0$ is
unaffected by zeroing $e$'s rewards, so the zeroed problem's supremum is at
least $\vfun_{e0}$.  For failure of equality on faithful models, the mechanism is that a
zero-reward activity can still relieve a constraint on other activities.
Take $\max\ \varepsilon x_1+x_2\ \text{s.t.}\ x_2\le x_1,\ x_1\le1,\
x\ge0$, with reward slot $\varepsilon>0$ on $e_1=x_1$; the probe zeroes
it, leaving $\max x_2$ with optimum $1$ at $(1,1)$,
while $\vfun_{e0}=0$; the zero-reward activity $x_1$ is still used, so
the would-be equality fails on a faithful model and no sound equality
test exists on the reward side; the rate-side probe
($a_{r1}\to M$ in $x_2\le x_1\le b$-type structure) does force $x_1\to0$ and
retains power.\Halmos

\subsection{Proof of Theorem \ref{thm:detect}}
Each case exhibits a specific test whose predicate the corrupted model
violates, given the certified precondition; soundness of the test for
faithful models makes the flag a correct conviction.

(i) A0 compares declared senses: flip is caught syntactically.
Behaviorally, objective probes are uninformative: for $x\ge0$, raising an
objective coefficient weakly raises minima and maxima alike, so the
flipped model satisfies A2's objective predicate and detection must come
from constraint-side probes, whose orientation is inherited from
feasible-set inclusion and reverses with the sense.  Let slot $j$ be a
rate slot in a $\le$-row (the $\ge$-row and ratio cases mirror).  Raising
$\theta_j$ shrinks the feasible set, so A2 requires the value to weakly
worsen in the asserted sense; the flipped model optimizes the opposite
sense over the same shrinking sets, so its reported value strictly
\emph{improves} in the asserted ordering whenever the probed row cuts the
flipped optimum, violating A2's weak inequality.  A6: the corrupted
(max-sense) value at the annihilation probe is $\ge$ the largest objective
of a feasible point and is strictly positive whenever any activity with
positive cost coefficient is feasible at level $>0$; the predicate requires
exactly $0$.

(ii) The corrupted model contains no occurrence of the crushed slot, so its
observations are constant along that slot's axis; at the certified crush
instance it reports the (finite) nominal status, while the predicate demands
infeasibility.

(iii) At a probe pair around a strictly binding instance of the flipped row,
the corrupted feasible set moves oppositely to the faithful one, producing a
strict improvement where A1 requires weak worsening (or a strict worsening
where A1 requires weak improvement); if the row carries a crush certificate,
the flipped row ($\ge0$ capacity read as $\ge$) is satisfied by any
nonnegative $x$ at the crush instance, so the corrupted model is feasible
there, violating A4; A6 similarly.

(iv) If the two swapped slots' qtypes assert opposite directions, A1 on
either slot observes the other row's response, whose orientation is
reversed at any strictly responsive probe.  If both are capacities but only
row $r$ has a crush certificate, the corrupted model at $r$'s crush instance
has $b_r=\theta^0_{r'}>0$ (the swapped value), and remains feasible whenever
the certificate was driven by row $r$ alone; the theorem's precondition that the swap breaks the certificate is exactly
this configuration.

(v) By Proposition~\ref{prop:limit} the \emph{predicate} compares
$\vfun'(M_2)$ with $\vfun'_{e0}$ on the corrupted model.  The corrupted model
lacks $e$'s cost slots, so its observations are constant in them:
$\vfun'(M)=\vfun'(\theta^0)$ for all $M$.  The certified precondition gives
$x_e=0$ feasible; the stated strictness condition $\vfun'(\theta^0)<
\vfun'_{e0}$ (using $e$ is strictly beneficial in the corrupted model when
its cost is deleted, which is generic whenever $e$ was used at all) is precisely the
violation of the equality predicate.

(vi) Rate misbinding sends $e_1$'s prohibitive-rate probe onto $e_2$'s
column: the corrupted model at $M$ squeezes $x_{e_2}\to0$ while leaving
$x_{e_1}$'s true consumption unpenalized, so
$\vfun'(M)\to\vfun'_{e_2 0}\ne\vfun'_{e_1 0}$ generically; the certified
predicate compares against $\vfun'_{e_1 0}$ and fails.  A7 detection under
congruence is Proposition~\ref{prop:reparam}(ii) applied to the swap
misfit.

(vii) A hard-coded rhs slot leaves observations constant along its axis, so
the crush instance retains the nominal (feasible) status against a certified
infeasibility (A4), and the A6 probe retains a strictly positive value; a
hard-coded rate slot leaves $x_e$ unsqueezed at the prohibitive probe,
reproducing case (v)'s violation.  Weak directional tests are satisfied
by constancy, which is the formal content of the requirement that
presence be established by limit probes.\Halmos

\subsection{Proof of Proposition \ref{prop:threshold}}
Consider the faithful family, indexed by $s\ge1$:
$\min\ x_1+x_2$ s.t.\ $x_1+x_2\ge 1$ (demand), $x_1\le s$ (capacity $A$),
$x_1+x_2\le 2s$ (capacity $B$, the audited row), $x\ge0$.
At the nominal instance the optimum is $1$.  The extreme probe multiplies
$b_B=2s$ by $0.001$; for $s\ge500$ the probed row still admits
$x_1+x_2\ge1$ (since $0.002s\ge1$), the optimum remains $1$, the change
ratio is $0<\tau$, and the tester flags the faithful model: false-positive
rate one on this family.  With $\tau=0$ no finite observation can be flagged
(any $r\ge0$ passes), so the tester is trivial.  The certified crush test
raises no false alarm here: at $b_B=0$ the declared structure is
infeasible, so the certificate holds and the test executes, and the
faithful candidate is likewise infeasible at that probe, so A4
\emph{passes} it, while a candidate that
omitted row $B$ remains feasible at $b_B=0$ and is convicted:
soundness and power coexist where the threshold rule cannot have
both.\Halmos

\subsection{Proof of Proposition \ref{prop:complexity}}
Immediate from the per-class solve counts in the statement, which read
off Table~\ref{tab:battery}'s probe definitions; A1 and A3 use
different probe offsets and relaxation modes, so their solves do not
pool.  Certificates add one declared-side solve per crush row and one
per entity, also $O(R+E)$; the all-pairs congruence scan contributes
the $O(E^2)$ worst-case term.\Halmos

\subsection{Soundness bound for the two-scale criterion (Remark \ref{rem:scale})}\label{ec:scale}
Fix a faithful candidate with certified precondition and write
$g(M)=\vfun_{e0}-\vfun(M)\ge0$ along the probe scale ($\sigma^\ast=\min$;
the mirrored case is identical).  By the proof of
Proposition~\ref{prop:limit}, $\vfun(M)$ is a pointwise minimum of
functions affine in $M$, hence concave and, by
Proposition~\ref{prop:direction}(ii), nondecreasing; so $g$ is
nonincreasing, convex, piecewise linear, and $g(M)=0$ for $M\ge\bar M$.
If $\bar M\le M_2$ the criterion accepts with gap $0$.  Otherwise
$g(M_2)>0$, and $g$ cannot be flat on $[M_1,M_2]$: a flat piece of a
convex nonincreasing function persists forever, contradicting
$g(\bar M)=0$.  Convexity bounds the slope beyond $M_2$ by the average
slope on $[M_1,M_2]$:
\[
g(M_2)\;=\;g(M_2)-g(\bar M)\;\le\;(\bar M-M_2)\,\lvert g'(M_2^+)\rvert
\;\le\;\frac{\bar M-M_2}{M_2-M_1}\,\bigl(g(M_1)-g(M_2)\bigr).
\]
The acceptance rule tolerates $g(M_2)\le 4\,(g(M_1)-g(M_2))$, which the
display guarantees whenever $\bar M-M_2\le 4(M_2-M_1)$, i.e.\
$\bar M\le 5M_2-4M_1$.\Halmos

\section{Additional Tables and Implementation Details}\label{ec:tables}

Table~\ref{ec:tab:parse} reports parser coverage,
Table~\ref{ec:tab:perclass} the distribution of convicting test classes
among deployment flags, Table~\ref{ec:tab:mcnemar} paired exact McNemar
tests for the selection policies, and Table~\ref{ec:tab:ci} Wilson
confidence intervals for headline rates; the remaining tables are cited
from the main text at their points of use.

Generation uses $N=8$ candidates (one greedy, seven at temperature
$0.8$, top-$p$ $0.95$, maximum $900$ new tokens); extraction is greedy;
all prompts ship verbatim in the replication package.  The same prompts
and sampling parameters drive both generators, the local
Qwen2.5-7B-Instruct and the DeepSeek \texttt{deepseek-chat} API
endpoint, whose server-reported model identifier was
\texttt{deepseek-v4-flash} throughout our access window (August
2026).  The API judge
uses the panel judges' prompt at temperature $0$ with $700$-token
responses; the self-consistency variant draws five samples at
temperature $0.8$ and takes the majority verdict.  The
solver is HiGHS 1.x via \texttt{highspy} with feasibility and
integrality tolerances $10^{-9}$ and the battery's one-sided violation
slack $\varepsilon=10^{-5}$ (relative); solver statuses that HiGHS
cannot separate (\texttt{kUnboundedOrInfeasible}) are treated as
inconclusive and never fire a test; experiments fix all random seeds
and run single-threaded per solve.

\input{ec_tables_gen}

\section{A Worked Audit Beyond the Benchmark Corpora}\label{ec:case}

The benchmark corpora contain no binary decisions, fixed charges, or
capacity-linking rows.  This section audits a depot-location
description with that structure (the classical capacitated
facility-location form; \citealt{balinski1965integer}), stated in full:

\begin{quote}\footnotesize
A regional bakery chain must decide which of three depots to operate:
Aston, Brook, and Carle.  Operating Aston costs \$840 per week and
provides capacity for 260 crates; operating Brook costs \$610 for 190
crates; operating Carle costs \$700 for 220 crates.  Four stores must be
supplied every week: store S1 requires 90 crates, store S2 requires 110
crates, store S3 requires 70 crates, and store S4 requires 100 crates.
Delivering one crate costs \$4 from Aston to S1, \$6 to S2, \$9 to S3,
and \$5 to S4; \$7 from Brook to S1, \$3 to S2, \$4 to S3, and \$8 to
S4; and \$6 from Carle to S1, \$5 to S2, \$7 to S3, and \$3 to S4.  A
depot can ship only if it is operated, and shipments from a depot cannot
exceed its capacity.  Which depots should be operated and how should
crates be shipped to minimize total weekly operating and delivery cost?
\end{quote}

The reference optimum is \$\numCaseOpt{} (operate Brook and Carle), and
the extractor recovers all 22 slots.  Of the $N{=}8$
DeepSeek-V4-flash candidates, \numCaseDSCorrect{} reproduces the
reference optimum with correct capacity-linking rows and passes both
operating points (\numCaseSolves{} solves); the remaining seven are
infeasible at the nominal data.  All
\numCaseQwenN{} Qwen2.5-7B candidates fail, unparseable, infeasible, or
at a degenerate \$0 optimum with requirements unbound.  Every organic
failure is execution-visible, and the battery stays silent on the
faithful candidate.  The threshold tester flags it with four warnings,
the first on the closed depot's capacity, whose crush probe changes
nothing (ratio $0<\tau_\ell$), the abundant-resource situation of
Proposition~\ref{prop:threshold}.

We inject three single errors into the faithful
candidate.  Transposing the Brook capacity (190) with
the S2 requirement (110), bindings moving with the numbers, yields
optimum \$\numCaseTransposeV{}; both operating points convict it with
two certificates that name the transposed pair:

\begin{quote}\footnotesize
\texttt{A1 rhs\_dir[s3]:\ v(dn)=3037\ \ v0=3780\ \ v(up)=inf\hfill(capacity)}\\
\texttt{A1 rhs\_dir[s7]:\ v(dn)=3780\ \ v0=3780\ \ v(up)=3370\hfill(requirement)}
\end{quote}

\noindent a capacity whose increase makes the model infeasible, a
requirement whose increase makes it cheaper.  Flipping the direction of
the Brook linking row, or deleting it outright, lowers the optimum to
\$\numCaseOmissionV{} and both operating points stay silent; the linking
coefficient enters the IR as a negated structural literal outside the
slot vocabulary, so the corruption is an omission in the sense of the
Deploy column of Table~\ref{tab:matrix} and is caught by execution
instead.  The case reproduces the division of labor of the aggregate
experiments: execution screens feasibility and value, the battery
certifies role structure, and the two are blind in complementary
places.

\end{document}

%% file: numbers.tex
\newcommand{\numSeedsTotal}{326}
\newcommand{\numCleanFPRpct}{0.0\%}
\newcommand{\numReloopFPRpct}{54.9\%}
\newcommand{\numCoreDetpct}{56.1\%}
\newcommand{\numCoreN}{3361}
\newcommand{\numMutantsTotal}{6947}

\newcommand{\numMSevenDetpct}{38.8\%}
\newcommand{\numMSevenAFiveResiduepct}{0.8\%}
\newcommand{\numAvgSolves}{25}
\newcommand{\numAvgBatterySec}{0.014}
\newcommand{\numBlindSetN}{5545}
\newcommand{\numBlindN}{1321}
\newcommand{\numBlindSharepct}{23.8\%}
\newcommand{\numBlindDetpct}{40.4\%}
\newcommand{\numCondDetpct}{70.0\%}
\newcommand{\numCondN}{1515}
\newcommand{\numUncondDetpct}{23.5\%}
\newcommand{\numUncondN}{2162}
\newcommand{\numMTwoCondDetpct}{100.0\%}
\newcommand{\numMTwoCondN}{270}

\newcommand{\numMFourCondDetpct}{60.5\%}

\newcommand{\numMSixCondDetpct}{48.4\%}

\newcommand{\numMSixUncondDetpct}{0.0\%}
\newcommand{\numMTenCondDetpct}{87.4\%}

\newcommand{\numMElevenCondDetpct}{62.4\%}

\newcommand{\numSynthDetFullpct}{78.0\%}
\newcommand{\numSynthDetFullN}{313}
\newcommand{\numSynthSolvable}{20}

\newcommand{\numBudgetNinetyFive}{88}
\newcommand{\numFiniteMFaithN}{24}
\newcommand{\numFiniteMMutN}{21}
\newcommand{\numFiniteMDetpct}{71.4\%}
\newcommand{\numAnnotAllpct}{44.5\%}
\newcommand{\numDeployAllpct}{29.2\%}
\newcommand{\numDeployMTwopct}{0.0\%}

\newcommand{\numDeployCleanFPRpct}{0.0\%}
\newcommand{\numAdjN}{155}
\newcommand{\numAdjTRV}{5}
\newcommand{\numAdjTRVpct}{3.2\%}

\newcommand{\numAdjExtOKpct}{76.8\%}

\newcommand{\numAdjExtBadpct}{20.0\%}
\newcommand{\numAdjDeservedpct}{23.2\%}
\newcommand{\numBenchNl}{239}
\newcommand{\numBenchME}{644}
\newcommand{\numBenchMC}{128}
\newcommand{\numBenchIor}{88}
\newcommand{\numSelMajNlpct}{39.3}
\newcommand{\numSelBatNlpct}{36.4}

\newcommand{\numSelOracleNlpct}{50.6}

\newcommand{\numSelGateNlpct}{41.0}
\newcommand{\numDSOracleNlpct}{87.7}

\newcommand{\numSelMajMEpct}{43.6}

\newcommand{\numSelOracleMEpct}{56.7}
\newcommand{\numSelReloopMEpct}{29.8}

\newcommand{\numDSOracleMEpct}{86.3}

\newcommand{\numSelOracleMCpct}{14.1}

\newcommand{\numSelOracleIorpct}{9.1}

\newcommand{\numGateSelWin}{5}
\newcommand{\numGateSelLose}{1}
\newcommand{\numGateSelP}{0.22}
\newcommand{\numReloopMEWin}{10}
\newcommand{\numReloopMELose}{99}
\newcommand{\numFlagPrec}{62.7\%}
\newcommand{\numFlagRec}{27.6\%}
\newcommand{\numGateFPNl}{13}

\newcommand{\numGatePrecNlpct}{92.0\%}
\newcommand{\numGateRecNlpct}{15.9\%}
\newcommand{\numVCFlagShareNl}{27.6\%}
\newcommand{\numVCFlaggedNl}{155}
\newcommand{\numVCCorrectNl}{562}
\newcommand{\numValidCands}{5769}
\newcommand{\numVCFlagSharePooled}{16.1\%}
\newcommand{\numVCFlaggedPooled}{316}
\newcommand{\numVCCorrectPooled}{1965}
\newcommand{\numGateVCFlagSharePooled}{7.2\%}
\newcommand{\numRhoStarNone}{1.41}
\newcommand{\numRhoStarReloop}{1.75}
\newcommand{\numBatFlagsPooled}{1094}
\newcommand{\numReloopFlagsPooled}{3508}
\newcommand{\numReloopFlagMult}{3.2}
\newcommand{\numOursF}{0.65}
\newcommand{\numOursDetpct}{47.7\%}
\newcommand{\numOursFPpct}{0.0\%}
\newcommand{\numReloopF}{0.77}
\newcommand{\numJudgeF}{0.82}
\newcommand{\numJudgeTwoF}{0.60}
\newcommand{\numOursJ}{0.48}
\newcommand{\numReloopJ}{0.22}
\newcommand{\numJudgeJ}{0.34}
\newcommand{\numJudgeTwoJ}{0.42}
\newcommand{\numReloopDetpct}{78.8\%}
\newcommand{\numReloopPanelFPpct}{56.7\%}
\newcommand{\numJudgeDetpct}{87.1\%}
\newcommand{\numJudgeFPpct}{53.3\%}
\newcommand{\numJudgeTwoDetpct}{43.2\%}
\newcommand{\numJudgeTwoFPpct}{1.7\%}
\newcommand{\numOursDetNCpct}{58.3\%}
\newcommand{\numOursJNC}{0.58}
\newcommand{\numReloopDetNCpct}{81.5\%}
\newcommand{\numReloopJNC}{0.25}
\newcommand{\numJudgeVThreeDetpct}{91.7\%}
\newcommand{\numJudgeVThreeFPpct}{20.0\%}
\newcommand{\numJudgeVThreeJ}{0.72}

\newcommand{\numJudgeVThreeDetNCpct}{97.2\%}
\newcommand{\numJudgeVThreeJNC}{0.77}

\newcommand{\numJudgeRznFPpct}{30.0\%}

\newcommand{\numJudgeSCDetpct}{93.2\%}
\newcommand{\numJudgeSCFPpct}{11.7\%}

\newcommand{\numJudgeSCJNC}{0.86}
\newcommand{\numPanelMutants}{132}
\newcommand{\numPanelMutantsNC}{108}
\newcommand{\numPanelClean}{60}
\newcommand{\numEscRecallNCpct}{98.1\%}
\newcommand{\numJBBoth}{62}
\newcommand{\numJBJudgeOnly}{43}
\newcommand{\numJBBatOnly}{1}
\newcommand{\numJudgeDeployPrecpct}{84.0\%}
\newcommand{\numJudgeDeployRecpct}{96.5\%}

\newcommand{\numJudgeDeployFPRpct}{35.6\%}
\newcommand{\numJudgeDeploySelNlpct}{43.9}
\newcommand{\numTriageSelNlpct}{43.9}

\newcommand{\numCaseDSCorrect}{1}

\newcommand{\numCaseQwenN}{8}
\newcommand{\numCaseSolves}{102}
\newcommand{\numCaseOpt}{2760}
\newcommand{\numCaseTransposeV}{3780}
\newcommand{\numCaseOmissionV}{2150}
\newcommand{\numParsedTotal}{351}
\newcommand{\numAnnTotal}{388}
\newcommand{\numParseCovpct}{90.5\%}

%% file: tables_gen.tex
\begin{table}[t]
\TABLE
{Detectability matrix on NL4OPT faithful seeds: annotated-mode and deployment-mode detection rates, ReLoop-style threshold tester, execution-blind shares, and per-class attribution of battery convictions.  Predictions from Theorem~\ref{thm:detect} and Propositions~\ref{prop:gauge}--\ref{prop:reparam}; M7--M9 are theory-invisible controls (M7: exchange-only).\label{tab:matrix}}
{\footnotesize\begin{tabular}{@{}llrrrrrl@{}}
\toprule
Op & Error class & $n$ & Battery & Deploy & Thresh. & Blind & Convicting classes \\
\midrule
M1 & direction flip & 849 & 73.3\% & 49.6\% & 63.6\% & 5\% & A1:50\%, A4:32\%, A5:25\% \\
M2 & omission & 846 & 42.9\% & 0.0\% & 86.8\% & 42\% & A4:33\%, A5:12\%, A7:2\% \\
M3 & rhs swap & 693 & 41.4\% & 16.5\% & 62.9\% & 12\% & A4:34\%, A6:23\%, A1:16\% \\
M4 & dropped obj.\ term & 647 & 44.2\% & 19.9\% & 89.8\% & 15\% & A7:31\%, A5:13\% \\
M5 & sense flip & 326 & 100.0\% & 100.0\% & 56.4\% & 1\% & A0:100\%, A6:55\%, A1:44\% \\
M6 & entity misbinding & 720 & 27.1\% & 26.8\% & 78.5\% & 33\% & A5:26\%, A7:2\% \\
M10 & hard-coded rhs & 709 & 60.8\% & 37.5\% & 91.3\% & 34\% & A6:34\%, A4:33\%, A7:4\% \\
M11 & hard-coded coef. & 755 & 38.8\% & 38.8\% & 94.8\% & 36\% & A7:32\%, A5:14\% \\
M7 & slot scale $\times10$ & 750 & 38.8\% & 38.5\% & 78.4\% & 30\% & A7:39\%, A5:1\% \\
M8 & objective $\times2$ & 326 & 0.0\% & 0.0\% & 54.9\% & 5\% & --- \\
M9 & redundant row & 326 & 0.0\% & 0.0\% & 54.9\% & 100\% & --- \\
\bottomrule
\end{tabular}}
{Threshold tester false-positive rate on the same faithful seeds: 54.9\% (battery, annotated and deployment mode alike: 0.0\%).  The Deploy column is the same mutant set verified from candidate-side assertions only (A4 skipped, ratio slots untestable); the M2 zero is structural, not statistical.}
\end{table}

\begin{table}[t]
\TABLE
{Best-of-8 selection accuracy (execution accuracy of the selected candidate) of the Qwen2.5-7B two-pass pipeline, all-problems denominator.\label{tab:select}}
{\footnotesize\begin{tabular}{@{}lrrrrrrrrr@{}}
\toprule
Benchmark & Greedy & Random & Majority & Thresh. & Battery & Gated & Weighted & Cascade & Oracle \\
\midrule
NL4OPT & 35.1 & 36.8 & 39.3 & 37.7 & 36.4 & 41.0 & 38.1 & 36.8 & 50.6 \\
MAMO EasyLP & 37.7 & 32.1 & 43.6 & 29.8 & 42.4 & 42.2 & 44.4 & 30.0 & 56.7 \\
MAMO ComplexLP & 11.7 & 10.2 & 12.5 & 10.9 & 12.5 & 12.5 & 12.5 & 10.9 & 14.1 \\
IndustryOR & 5.7 & 5.7 & 6.8 & 4.5 & 3.4 & 5.7 & 6.8 & 4.5 & 9.1 \\
\bottomrule
\end{tabular}}
{Accuracies in percent; labels are used only for scoring, never by any policy.  Gated = hard filtering by the binding-consistent deployment battery of Section~\ref{sec:gate}.}
\end{table}

\begin{figure}[t]
\FIGURE
{\includegraphics[width=.46\textwidth]{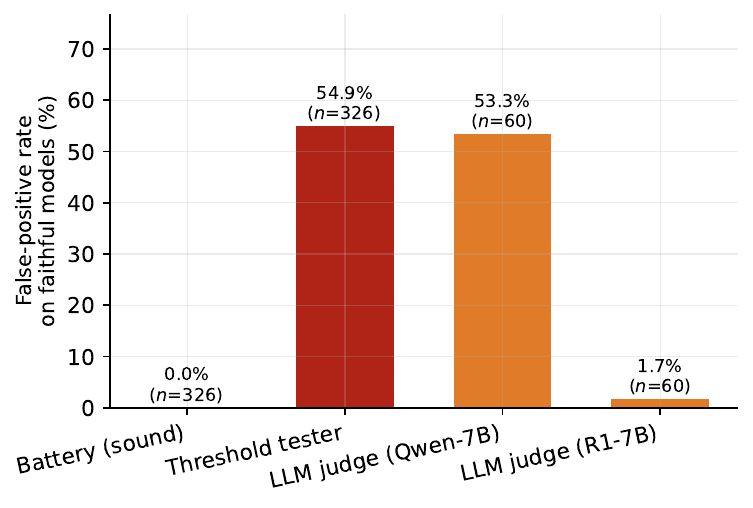}\quad\includegraphics[width=.5\textwidth]{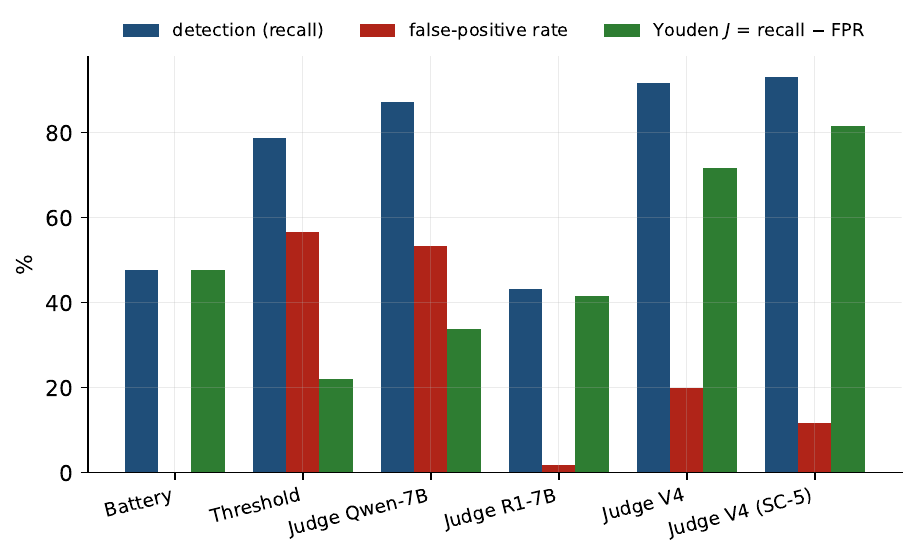}}
{Soundness in practice.  Left: false-positive rates on faithful models.  Right: detection, false-positive rate, and Youden $J$ on the stratified panel.\label{fig:fpr}}
{}
\end{figure}

\begin{figure}[t]
\FIGURE
{\includegraphics[width=.9\textwidth]{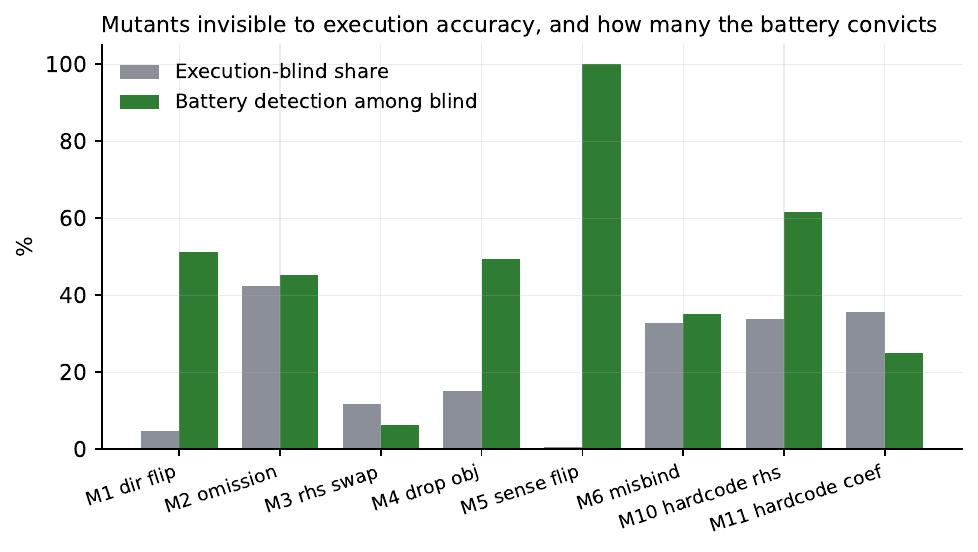}}
{Execution-blind mutants (value-coincident at the nominal instance) and battery detection among them.\label{fig:blind}}
{}
\end{figure}

\begin{figure}[t]
\FIGURE
{\includegraphics[width=.48\textwidth]{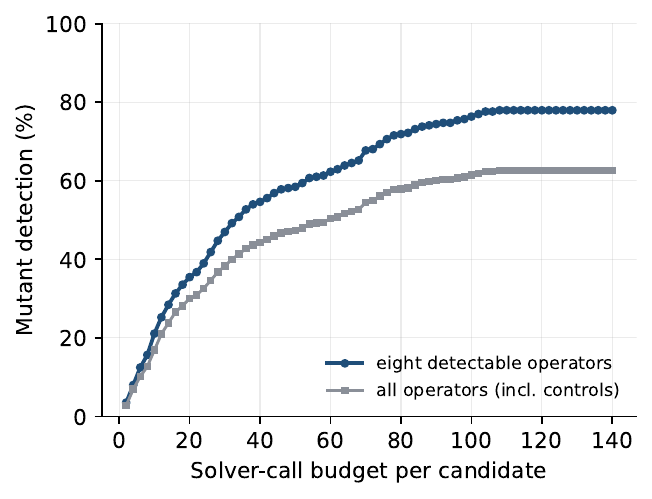}\includegraphics[width=.48\textwidth]{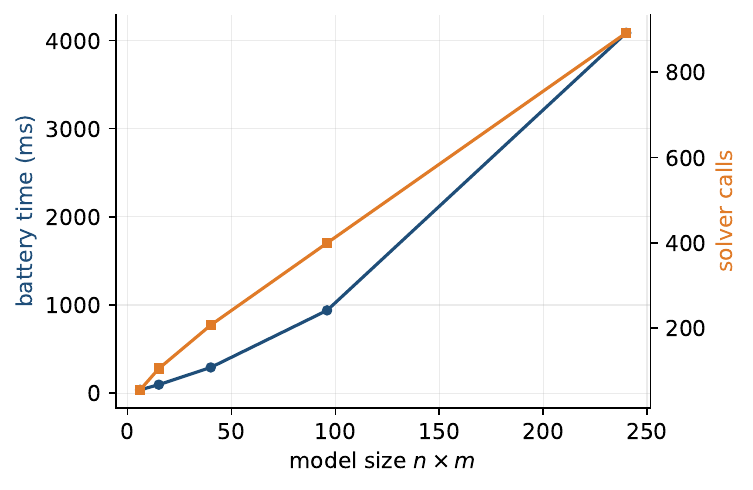}}
{Left: detection power of the battery versus solver-call budget on the certified synthetic family, on the eight detectable operators and on all operators including the invisibility controls.  Right: battery cost versus model size.\label{fig:budget}}
{}
\end{figure}

%% file: ec_tables_gen.tex
\begin{table}[h]
\TABLE{NL4OPT annotation parser coverage (the five most frequent failure reasons are itemized).\label{ec:tab:parse}}
{\begin{tabular}{@{}lr@{}}\toprule
Parsed into IR & 351 \\
Unparseable annotations & 37 \\
\quad ratio needs var + percentage & 5 \\
\quad var 'premium one' unknown & 1 \\
\quad var 'premium batch' unknown & 1 \\
\quad var 'single-load machine' unknown & 1 \\
\quad var 'electric-based stores' unknown & 1 \\
\bottomrule\end{tabular}}{}
\end{table}

\begin{table}[h]
\TABLE{Stronger-generator study: best-of-8 selection accuracy with DeepSeek-V4-flash as the two-pass generator (same prompts, $N{=}8$, all-problems denominator).\label{ec:tab:deepseek}}
{\footnotesize\begin{tabular}{@{}lrrrrrr@{}}\toprule
Benchmark & Greedy & Majority & Thresh. & Battery & Gated & Oracle \\
\midrule
NL4OPT & 82.4 & 83.2 & 83.2 & 83.2 & 83.2 & 87.7 \\
MAMO EasyLP & 83.6 & 84.7 & 83.1 & 84.8 & 84.8 & 86.3 \\
MAMO ComplexLP & 21.1 & 22.1 & 20.0 & 23.2 & 23.2 & 24.2 \\
IndustryOR & 25.9 & 27.1 & 24.7 & 22.4 & 22.4 & 32.9 \\
\bottomrule\end{tabular}}{}
\end{table}

\begin{table}[h]
\TABLE{Adjudication of all 155 value-correct-but-flagged NL4OPT certificates (LLM first pass, human-audited; per-case records in the replication package).\label{ec:tab:adjudication}}
{\begin{tabular}{@{}lrr@{}}\toprule
Category & Cases & Share \\
\midrule
Certificate exposes a true role violation & 5 & 3.2\% \\
Assertion-side (extraction) artifact, candidate also structurally wrong & 31 & 20.0\% \\
Assertion-side (extraction) artifact, candidate faithful to the text & 119 & 76.8\% \\
\bottomrule\end{tabular}}
{Dominant artifact mechanisms: qtype confusion (per-unit rates typed as requirement/capacity), objective-typed cost slots that belong to budget rows, and entity-label granularity breaking exchange pairing.}
\end{table}

\begin{table}[h]
\TABLE{Distribution of failed test classes among deployment-mode battery flags, pooled over the four Qwen pools (per-benchmark distributions in the artifacts).\label{ec:tab:perclass}}
{\begin{tabular}{@{}lrr@{}}\toprule
Class & All flags & Among value-correct flags \\
\midrule
A1 & 619 & 144 \\
A2 & 297 & 55 \\
A3 & 217 & 97 \\
A7 & 112 & 49 \\
A5 & 23 & 0 \\
\bottomrule\end{tabular}}{}
\end{table}

\begin{table}[h]
\TABLE{Two-scale limit criterion across four decades of $M$: faithful pass rates and misbinding detection on the certified synthetic family.\label{ec:tab:finiteM}}
{\begin{tabular}{@{}lrr@{}}\toprule
$(M_1,M_2)$ & Faithful pass & M6 detected \\
\midrule
$(10^{3},10^{6})$ & 100\% & 71\% \\
$(10^{4},10^{7})$ & 100\% & 71\% \\
$(10^{6},10^{9})$ & 100\% & 71\% \\
$(10^{7},10^{10})$ & 100\% & 71\% \\
\bottomrule\end{tabular}}{}
\end{table}

\begin{table}[h]
\TABLE{Benchmark label audit.  Annotation-vs-label cases are certified by solving the official annotation both as an LP and as an all-integer program (both differ from the label); the accompanying notes are generated by a mechanical rule, not by human review.\label{ec:tab:audit}}
{\footnotesize\begin{tabular}{@{}lrrl@{}}\toprule
Benchmark & Cases & Confirmed defects & Notes \\
\midrule
nl4opt annotation vs label & 18 & 18 & LP and integer IR both differ; 189 agree (71 integer-only) \\
nl4opt consensus & 19 & 0 & battery-passing 6/8 consensus differs from label \\
mamo\_easy consensus & 15 & 0 & battery-passing 6/8 consensus differs from label \\
mamo\_complex consensus & 3 & 0 & battery-passing 6/8 consensus differs from label \\
industryor consensus & 1 & 0 & battery-passing 6/8 consensus differs from label \\
\bottomrule\end{tabular}}{}
\end{table}

\begin{table}[h]
\TABLE{The 18 NL4OPT annotation-vs-label mismatches.  \emph{ann LP}/\emph{ann int}: optimum of the official annotation solved as LP / all-integer; notes are mechanical, not human adjudications.\label{ec:tab:auditdetail}}
{\scriptsize\begin{tabular}{@{}lrrr@{}}\toprule
Problem & ann LP & ann int & label \\
\midrule
2005236115 & 2333 & 2333 & 2400 \\
1703643437 & 47.78 & 50 & -1e+05 \\
-642253022 & 6 & 6 & 8 \\
-2027758451 & 16.67 & 17 & 18 \\
1165597365 & 7.5 & -- & -1e+05 \\
-272035411 & 3 & 3 & 2.3 \\
1637604355 & 2.4 & 2.4 & 4 \\
1402521519 & 1.125e+04 & 1.128e+04 & 2500 \\
-1765797791 & 55.56 & 54 & 50 \\
2041857060 & 5625 & 5625 & 7500 \\
-1819716628 & 21.33 & 20.8 & 6.4 \\
615752596 & 85.71 & 80 & -1e+05 \\
-1062928915 & 0.72 & 0.5 & 6.5e+04 \\
-114149479 & 1200 & 1200 & 648 \\
358766780 & 45.45 & 46 & 48 \\
-169306566 & 56 & 56 & 32 \\
-1243268146 & 3200 & 3200 & -1e+05 \\
-527722703 & 1.239e+04 & 1.26e+04 & 1.286e+04 \\
\bottomrule\end{tabular}}{}
\end{table}

\begin{table}[h]
\TABLE{Scale-robustness boundary of the implemented two-scale criterion (Remark~\ref{rem:scale}): faithful two-cost family $\min x_1 + Kx_2$, probe scales $(M_1,M_2)=(10^6,10^9)$, soundness bound $\bar M\le 5M_2-4M_1$.\label{ec:tab:scalebound}}
{\footnotesize\begin{tabular}{@{}rrccc@{}}\toprule
$K$ & $\bar M$ & Predicted pass & Observed pass & Enlarged scales \\
\midrule
$10^{6}$ & 5e+05 & yes & yes & yes \\
$10^{8}$ & 5e+07 & yes & yes & yes \\
$10^{9}$ & 5e+08 & yes & yes & yes \\
$10^{9}$ & 2.5e+09 & yes & yes & yes \\
$10^{9}$ & 4.5e+09 & yes & yes & yes \\
$10^{9}$ & 4.95e+09 & yes & yes & yes \\
$10^{10}$ & 5.25e+09 & no & no & yes \\
$10^{10}$ & 1e+10 & no & no & yes \\
$10^{11}$ & 5e+10 & no & no & yes \\
$10^{12}$ & 5e+11 & no & no & yes \\
\bottomrule\end{tabular}}{}
\end{table}

\begin{table}[h]
\TABLE{Two-pass pipeline yield per benchmark (Qwen generator): problems entering evaluation, problems with $\ge$1 valid candidate, and candidate-level validity and battery pass counts.\label{ec:tab:yield}}
{\footnotesize\begin{tabular}{@{}lrrrrr@{}}\toprule
Benchmark & Problems & $\ge$1 valid & Candidates & Valid & Battery pass \\
\midrule
NL4OPT & 239 & 236 & 1912 & 1474 & 1058 \\
MAMO EasyLP & 644 & 626 & 5152 & 3612 & 3116 \\
MAMO ComplexLP & 128 & 79 & 1024 & 370 & 275 \\
IndustryOR & 88 & 68 & 704 & 313 & 226 \\
\bottomrule\end{tabular}}{}
\end{table}

\begin{table}[h]
\TABLE{Paired exact McNemar tests between selection policies (all-problems denominator): wins/losses of the row policy against plain majority voting.\label{ec:tab:mcnemar}}
{\footnotesize\begin{tabular}{@{}llrrr@{}}\toprule
Benchmark & Policy & Wins & Losses & $p$ \\
\midrule
NL4OPT & battery & 6 & 13 & 0.1671 \\
NL4OPT & battery (gated) & 5 & 1 & 0.2188 \\
NL4OPT & weighted & 3 & 6 & 0.5078 \\
NL4OPT & threshold & 11 & 15 & 0.5572 \\
NL4OPT & cascade & 11 & 17 & 0.3449 \\
\addlinespace
MAMO EasyLP & battery & 8 & 16 & 0.1516 \\
MAMO EasyLP & battery (gated) & 6 & 15 & 0.0784 \\
MAMO EasyLP & weighted & 7 & 2 & 0.1797 \\
MAMO EasyLP & threshold & 10 & 99 & 0.0000 \\
MAMO EasyLP & cascade & 11 & 99 & 0.0000 \\
\addlinespace
MAMO ComplexLP & battery & 0 & 0 & 1.0000 \\
MAMO ComplexLP & battery (gated) & 0 & 0 & 1.0000 \\
MAMO ComplexLP & weighted & 0 & 0 & 1.0000 \\
MAMO ComplexLP & threshold & 0 & 2 & 0.5000 \\
MAMO ComplexLP & cascade & 0 & 2 & 0.5000 \\
\addlinespace
IndustryOR & battery & 0 & 3 & 0.2500 \\
IndustryOR & battery (gated) & 0 & 1 & 1.0000 \\
IndustryOR & weighted & 0 & 0 & 1.0000 \\
IndustryOR & threshold & 1 & 3 & 0.6250 \\
IndustryOR & cascade & 1 & 3 & 0.6250 \\
\addlinespace
\bottomrule\end{tabular}}{}
\end{table}

\begin{table}[h]
\TABLE{Wilson 95\% confidence intervals for headline rates.\label{ec:tab:ci}}
{\begin{tabular}{@{}lrrl@{}}\toprule
Rate & $k$ & $n$ & 95\% CI \\
\midrule
Clean FPR (battery) & 0 & 326 & [0.0\%, 1.2\%] \\
Core detection M1--M5 & 1884 & 3361 & [54.4\%, 57.7\%] \\
Detection among execution-blind & 534 & 1321 & [37.8\%, 43.1\%] \\
True violations among adjudicated flags & 5 & 155 & [1.4\%, 7.3\%] \\
\bottomrule\end{tabular}}{}
\end{table}